\documentclass[10pt,onecolumn,aps,prd,preprintnumbers,showpacs,superscriptaddress,nofootinbib,amsmath,amssymb,floats,floatfix,showkeys,notitlepage,longbibliography]{revtex4-2}
\usepackage{orcidlink}
\usepackage{comment}
\usepackage{lipsum}
\usepackage{graphicx}
\usepackage{subfigure}
\usepackage{palatino}
\usepackage{sans}
\usepackage{array}
\usepackage{physics}
\usepackage{hyperref}
\hypersetup{colorlinks=true,linkcolor=blue,urlcolor=blue,citecolor=blue}
\usepackage[toc,page]{appendix}
\usepackage[normalem]{ulem}
\usepackage{adjustbox}
\usepackage{latexsym}
\usepackage{amsmath}
\usepackage{amssymb}
\usepackage{amsfonts}
\numberwithin{equation}{section}
\usepackage{mathrsfs}
\usepackage{dcolumn}
\usepackage{bm}
\usepackage{tikz}
\usetikzlibrary{decorations.pathmorphing}
\usepackage{bigints}
\usepackage{array,tabularx,multirow,booktabs}
\usepackage[tracking=true]{microtype}
\usepackage{soul} 
\SetTracking{}{500}
\SetTracking{encoding={*}, shape=sc}{40}
\UseRawInputEncoding 
\allowdisplaybreaks
\usepackage{mathrsfs}
\usepackage[utf8]{inputenc}
\usepackage{xcolor} 

\begin{document} \sloppy

\title{Maxwell-Jüttner averaged nonaffine theory of relativistic shear viscosity}

\author{Reggie C. Pantig \orcidlink{0000-0002-3101-8591}} 
\email{rcpantig@mapua.edu.ph}
\affiliation{Physics Department, School of Foundational Studies and Education, Map\'ua University, 658 Muralla St., Intramuros, Manila 1002  Philippines.}

\author{Ali \"Ovg\"un \orcidlink{0000-0002-9889-342X}}
\email{ali.ovgun@emu.edu.tr}
\affiliation{Physics Department, Eastern Mediterranean University, Famagusta, 99628 North
Cyprus via Mersin 10, Turkiye.}

\begin{abstract}
We formulate a Maxwell-Jüttner averaged extension of relativistic nonaffine shear-viscosity theory for a classical gas of massive particles. The starting point is the relativistic nonaffine response formula derived from generalized Langevin dynamics, where proper-momentum dissipation produces a Lorentz-factor enhancement of the viscosity. Since a thermal relativistic gas is not characterized by a single Lorentz factor, we replace this scalar enhancement by an equilibrium momentum average over the Maxwell-Jüttner distribution.  Within the scalar transverse mean-field closure used in this work, the minimal relativistic correction is the Maxwell-J\"uttner mean of the Lorentz factor and becomes a finite-mass enhancement function of $\zeta=mc^2/(k_BT)$. In the general closure, the averaged object is the full product of the Lorentz factor, the zero-frequency memory kernel, and the nonaffine affine-force correlator. This construction separates kinematic relativistic enhancement from dynamical dissipation encoded in the bath memory and collective mode spectrum. We derive the nonrelativistic and ultrarelativistic limits,  show that a specified effective hard-sphere scaling reproduces the classical square-root law as a consistency test, and identify the assumptions under which a cubic high-temperature law follows. Comparison with relativistic kinetic-theory examples clarifies that the memory kernel is not equivalent to a relaxation time unless a microscopic closure is specified.  Retaining the full frequency dependence instead yields a relaxation spectrum. A single dominant pole reduces this spectrum to one shear relaxation time and gives the usual transient shear equation.
\end{abstract}

\keywords{relativistic shear viscosity;
Maxwell-Jüttner distribution;
nonaffine response theory;
generalized Langevin dynamics;
memory kernels;
relativistic hydrodynamics}

\maketitle

\section{Introduction}
\label{sec_introduction}

Shear viscosity governs the irreversible relaxation of momentum anisotropies in relativistic many-body systems and enters the dissipative part of the energy-momentum tensor \cite{Kubo_1957,Anderson:1974nyl,Denicol:2012cn,Zaccone_2023}. 
Kinetic theory, Green-Kubo response, relativistic stochastic dynamics, and hydrodynamic formulations describe this dissipation with different microscopic variables and approximations \cite{Kubo_1957,Anderson:1974nyl,Dunkel:2008ngc,Debbasch_1997,Rocha:2021zcw,Mori:1965oqj,Zwanzig:1961zz}. 
For relativistic gases, the equilibrium measure, collision physics, relaxation scales, and hydrodynamic frame all affect the transport description \cite{Dunkel:2006nk,Kovtun:2019hdm,Bemfica:2019knx,Ambrus:2023qcl}.

Relativistic nonaffine response theory provides a complementary microscopic formulation based on generalized Langevin dynamics and collective shear modes \cite{Petrosyan:2021lqi,Zadra:2023ceq,Zaccone_2024}. 
Its dissipative response is expressed through a collective-mode density, an affine-force correlator, and a friction memory response \cite{Lemaitre_2006,Zaccone_2011,Zaccone_2023}. 
The relativistic extension introduces a scalar Lorentz-factor enhancement of the momentum-loss channel \cite{Petrosyan:2021lqi,Zadra:2023ceq,Zaccone_2024}.

A thermal relativistic gas does not have a single Lorentz factor. 
For a classical gas in equilibrium, the particle momenta follow the Maxwell-J\"uttner distribution and $\gamma$ fluctuates across the ensemble \cite{Juttner_1911,Dunkel:2006nk,Dunkel:2008ngc}. 
We therefore replace the fixed scalar enhancement by a thermal average within the scalar transverse mean-field branch and state explicitly in Sec.~\ref{sec_rel_nonaffine} how this branch differs from the full tensorial many-particle response. 
We use the inverse relativistic temperature $\zeta=mc^2/(k_BT)$ and the equivalent dimensionless temperature $\theta=k_BT/(mc^2)=\zeta^{-1}$. 
The general thermal average also permits momentum dependence in the friction response and affine-force correlator.

Our formulation does not replace the relativistic Boltzmann equation. 
It gives a finite-mass response representation whose assumptions can be tested against kinetic theory and explicit finite-memory models \cite{Anderson:1974nyl,Rocha:2021zcw,Denicol:2019lio,Ambrus:2023qcl}. 
Microscopic memory and hydrodynamic relaxation scales need not coincide. 
Projection-operator and nonequilibrium-statistical approaches distinguish a memory response generated by unresolved degrees of freedom from relaxation parameters that emerge after a hydrodynamic reduction \cite{Mori:1965oqj,Zwanzig:1961zz,Denicol:2011fa,Harutyunyan:2018cmm}. 
Hydrodynamic use of the resulting shear coefficient further requires a causal and stable constitutive theory with a specified frame. 
Transient and second-order formulations introduce additional relaxation structure \cite{Israel:1979wp,Baier:2007ix,Denicol:2012cn}, while the instability of standard first-order formulations and the development of stable relativistic alternatives motivate careful treatment of causality and frame choice \cite{Hiscock:1985zz,Kovtun:2019hdm,Bemfica:2019knx,Florkowski:2017olj}.

In this paper, we formulate the Maxwell-J\"uttner averaged nonaffine viscosity functional, derive its limiting forms, and identify the assumptions under which it reduces to known nonrelativistic and ultrarelativistic scalings. 
Section \ref{sec_rel_nonaffine} reviews the relativistic nonaffine viscosity formula and isolates the unaveraged Lorentz enhancement. 
Section \ref{sec_juttner_average} develops the Maxwell-J\"uttner equilibrium averages needed for the finite-$\zeta$ theory. 
Section \ref{sec_averaged_viscosity} derives the averaged viscosity functional and the associated enhancement function. 
Section \ref{sec_limits} analyzes the nonrelativistic, moderately relativistic, and ultrarelativistic limits. 
Section \ref{sec_kinetic_examples} compares the result with relativistic kinetic-theory examples, and Sec. \ref{sec_hydro_consistency} discusses hydrodynamic consistency and memory-induced relaxation. 
We use the metric signature $(+,-,-,-)$, reserve $\gamma$ exclusively for the Lorentz factor, denote shear strain by $\varepsilon_{xy}$, and keep $c$, $k_B$, and $\hbar$ explicit unless otherwise stated.

\section{Relativistic Nonaffine Viscosity and Its Unaveraged Limit}
\label{sec_rel_nonaffine}
\subsection{Relativistic Langevin origin of the nonaffine response}
\label{subsec_rel_langevin_nonaffine}
We begin with the relativistic generalized Langevin equation obtained from a particle-bath construction of Caldeira-Leggett type \cite{Caldeira:1982iu,Petrosyan:2021lqi,Zadra:2023ceq}. 
The bath variables represent unresolved microscopic degrees of freedom, and their elimination produces a memory kernel rather than an instantaneous friction coefficient, as in the standard generalized Langevin construction \cite{Zwanzig_1973,Dunkel:2008ngc}. 
For a tagged particle with four-momentum $p^\mu$, the reduced equation may be written in the laboratory time parameter as
\begin{equation}
\frac{d p^\mu}{dt}
=
F^\mu_{\rm ext}
+
F^\mu_{\rm fl}
-
\frac{1}{m}
\int_{0}^{t}
K^\mu{}_{\nu}(t,t')p^\nu(t')\,dt' .
\label{2.1}
\end{equation}
Here $F^\mu_{\rm ext}$ denotes the conservative external force, $F^\mu_{\rm fl}$ is the fluctuating bath force, and $K^\mu{}_{\nu}$ is the relativistic friction-memory tensor. 
Equation \eqref{2.1} is not yet a viscosity formula; it is the microscopic dynamical input from which the dissipative response is extracted. 
The stochastic force has zero mean in linear response, whereas its second cumulant is constrained by a fluctuation-dissipation relation determined by the same bath kernel \cite{Zwanzig_1973,Dunkel:2008ngc,Petrosyan:2021lqi,Zadra:2023ceq}. 
Since the present section concerns the deterministic shear response, we work with the mean equation after the noise average has been taken.

Let $s_a$ denote the nonaffine displacement of particle component $a=(i,\alpha)$ away from the position prescribed by the imposed affine shear deformation. 
The linearized response equation used below is the massful form of the relativistic nonaffine Langevin equation, with the memory kernel convention chosen so that $\nu$ has dimensions of mass over time, consistently with the corrected nonaffine-viscosity convention \cite{Zaccone_2023,Zaccone_2024b}. 
For a homogeneous shear strain $\varepsilon_{xy}(t)$, the deterministic linearized equation reads
\begin{equation}
m\gamma\,\ddot{s}_{a}(t)
+
\gamma\int_{-\infty}^{t}
\nu(t-t')\dot{s}_{a}(t')\,dt'
+
\sum_{b}H_{ab}s_b(t)
=
\Xi_{a,xy}\varepsilon_{xy}(t) .
\label{2.2}
\end{equation}

The factor $m\gamma$ in Eq. \eqref{2.2} is the scalar transverse closure inherited from the relativistic nonaffine formulation \cite{Petrosyan:2021lqi,Zadra:2023ceq,Zaccone_2024}. 
We do not identify it with the exact relativistic inertia of a thermal many-particle system. 
For particle $i$, linearizing $p_{i\alpha}=m\gamma_i v_{i\alpha}$ with respect to a velocity perturbation gives
\begin{equation*}
\delta p_{i\alpha}
=
\sum_\beta
\mathsf M^{(i)}_{\alpha\beta}\,\delta v_{i\beta},
\qquad
\mathsf M^{(i)}_{\alpha\beta}
=
m\left(
\gamma_i\delta_{\alpha\beta}
+
\frac{\gamma_i^3}{c^2}v_{i\alpha}v_{i\beta}
\right).
\tag{II.2a}
\label{2.tensor.inertia}
\end{equation*}
This relation follows directly from relativistic momentum and introduces no further dynamical assumption. 
At fixed speed, isotropy gives $\langle v_{i\alpha}v_{i\beta}\rangle_\Omega=v_i^2\delta_{\alpha\beta}/3$ and $v_i^2/c^2=1-\gamma_i^{-2}$. 
The angularly averaged inertia is therefore
\begin{equation*}
\left\langle
\mathsf M^{(i)}_{\alpha\beta}
\right\rangle_\Omega
=
m\,\mu(\gamma_i)\delta_{\alpha\beta},
\qquad
\mu(\gamma)
=
\frac{\gamma^3+2\gamma}{3}.
\tag{II.2b}
\label{2.isotropic.inertia}
\end{equation*}
Thus isotropy makes the one-particle inertia scalar, but the scalar factor is $\mu(\gamma)$ rather than $\gamma$. 
The inertial part of Eq. \eqref{2.2} follows exactly for a perturbation transverse to the particle momentum, $v_i^\alpha\delta v_{i\alpha}=0$, because the second term in Eq. \eqref{2.tensor.inertia} then vanishes. 
A thermal shear perturbation does not impose this transversality particle by particle.

At a fixed instantaneous set of particle momenta, the corresponding many-particle linear response has the form
\begin{equation*}
\sum_b \mathsf M_{ab}\ddot s_b(t)
+
\sum_b\int_{-\infty}^{t}
\mathsf D_{ab}(t-t')\dot s_b(t')\,dt'
+
\sum_b H_{ab}s_b(t)
=
\Xi_{a,xy}\varepsilon_{xy}(t),
\tag{II.2c}
\label{2.manyparticle}
\end{equation*}
where $\mathsf M_{ab}=\delta_{ij}\mathsf M^{(i)}_{\alpha\beta}$ for $a=(i,\alpha)$ and $b=(j,\beta)$, while $\mathsf D_{ab}$ denotes the retarded friction matrix. 
We evaluate Eq. \eqref{2.manyparticle} about an instantaneous mechanical reference configuration. 
The modal projection below assumes that its Hessian changes slowly over the linear response interval. 
If the reference configuration evolves on the same time scale as the imposed shear, then $H_{ab}(t)$ and its eigenvectors become time dependent and the changing basis generates additional mode-coupling terms. 
That regime lies outside the present scalar modal theory.
Projection onto the Hessian eigenvectors gives
\begin{equation*}
\mathsf M_{pq}
=
\sum_{ab}e_a^{(p)}\mathsf M_{ab}e_b^{(q)},
\qquad
\mathsf D_{pq}(t)
=
\sum_{ab}e_a^{(p)}\mathsf D_{ab}(t)e_b^{(q)},
\tag{II.2d}
\label{2.projected.matrices}
\end{equation*}
and hence
\begin{equation*}
\sum_q\left[
\mathsf M_{pq}\ddot s_q(t)
+
\int_{-\infty}^{t}
\mathsf D_{pq}(t-t')\dot s_q(t')\,dt'
\right]
+
m\omega_p^2s_p(t)
=
\Xi_{p,xy}\varepsilon_{xy}(t).
\tag{II.2e}
\label{2.coupled.modes}
\end{equation*}
Different particle Lorentz factors can therefore generate off-diagonal mode couplings. 
Even after the angular average in Eq. \eqref{2.isotropic.inertia}, one obtains
\begin{equation*}
\left\langle\mathsf M_{pq}\right\rangle_\Omega
=
m\sum_i\mu(\gamma_i)
\sum_\alpha e_{i\alpha}^{(p)}e_{i\alpha}^{(q)},
\tag{II.2f}
\label{2.angular.projected}
\end{equation*}
which is not diagonal for a generic set of particle speeds. 
If the equilibrium momenta are identically distributed and statistically independent of the collective eigenvectors, the subsequent Maxwell-J\"uttner average gives
\begin{equation*}
\left\langle\mathsf M_{pq}\right\rangle_{J,\Omega}
=
m\left\langle\mu(\gamma)\right\rangle_J\delta_{pq}.
\tag{II.2g}
\label{2.thermal.projected}
\end{equation*}
The same diagonal reduction of the retarded friction matrix requires the analogous absence of persistent correlations between particle momenta and collective eigenvectors. 
These relations show that angular and equilibrium averages can restore a scalar ensemble response under stated statistical assumptions, while they do not derive the factor $m\langle\gamma\rangle_J$ used by the minimal theory.

In the remainder of the paper, we therefore average the already diagonalized fixed-$\gamma$ response in Eq. \eqref{2.2}. 
We treat this step as an additional scalar transverse mean-field approximation. 
It is appropriate when the microscopic shear response is predominantly transverse, momentum directions decorrelate from collective eigenvectors, and the off-diagonal projected inertia and friction elements remain small relative to the diagonal modal terms. 
At leading nonrelativistic order, $\gamma_i\to1$ and $\mu(\gamma_i)\to1$, so the tensorial and scalar inertial descriptions coincide. 
When these conditions fail, thermal fluctuations can mix the Hessian modes and Eq. \eqref{2.coupled.modes} must replace the scalar modal response used below.

The Hessian and the affine-force field are defined by
\begin{equation}
H_{ab}
=
\left.
\frac{\partial^2 U}{\partial r_a\partial r_b}
\right|_{\varepsilon=0},
\qquad
\Xi_{a,xy}
=
\left.
\frac{\partial f_a}{\partial \varepsilon_{xy}}
\right|_{\varepsilon=0}
=
-
\left.
\frac{\partial^2 U}{\partial r_a\partial\varepsilon_{xy}}
\right|_{\varepsilon=0}.
\label{2.3}
\end{equation}
Thus $\Xi_{a,xy}$ is not an external force inserted by hand; it is the force generated when a particle is first placed at its affine position under shear. 
The nonaffine displacement is the additional relaxation required to restore mechanical equilibrium, and it is this relaxation channel that produces the mode-resolved dissipative response \cite{Lemaitre_2006,Zaccone_2011,Zaccone_2023}.

\subsection{Mode-resolved nonaffine viscosity formula}
\label{subsec_mode_resolved_nonaffine_viscosity}
We diagonalize the Hessian by mass-normalized normal modes $e_a^{(p)}$ according to
\begin{equation}
\sum_b H_{ab}e_b^{(p)}
=
m\omega_p^2 e_a^{(p)},
\qquad
\sum_a e_a^{(p)}e_a^{(q)}
=
\delta_{pq}.
\label{2.4}
\end{equation}

For the real-frequency sums and integrals used below, the index $p$ labels only the mechanically stable subspace of the reference Hessian after translational zero modes have been removed.
In a fluid, this reference configuration is not a permanent solid.
We use either a quenched inherent structure associated with the instantaneous configuration or a short-lived mechanically stable local configuration that remains approximately fixed over the linear response interval, as in nonaffine normal-mode formulations \cite{Lemaitre_2006,Zaccone_2023}.
This requires structural evolution to be slower than the response carried by the retained modes.
If $H$ has a negative eigenvalue $\lambda_p<0$, then $\omega_p$ is imaginary and that direction does not enter the real-frequency density $g(\omega_p)$.
A strongly overdamped excitation likewise cannot be represented by the Hermitian Hessian eigenproblem in Eq. \eqref{2.4}.
It requires a generalized damped spectral problem in which inertia and dissipation enter the eigenvalue equation together.
These sectors lie outside the real-frequency formula used below.
The direct normal-mode interpretation is therefore aimed at dense or correlated fluids with transient local structure.
For a very dilute gas with no quasi-stable reference spectrum, kinetic theory gives the appropriate microscopic description, while the Debye reduction should be read only as an effective scaling model.

The displacement and affine force are expanded as $s_a(t)=\sum_p e_a^{(p)}s_p(t)$ and $\Xi_{p,xy}=\sum_a e_a^{(p)}\Xi_{a,xy}$. 
 Substituting these decompositions into Eq. \eqref{2.2} gives the diagonal linear modal equation within the scalar transverse mean-field approximation
\begin{equation}
m\gamma\,\ddot{s}_{p}(t)
+
\gamma\int_{-\infty}^{t}
\nu(t-t')\dot{s}_{p}(t')\,dt'
+
m\omega_p^2 s_p(t)
=
\Xi_{p,xy}\varepsilon_{xy}(t).
\label{2.5}
\end{equation}
We use the Fourier convention $A(t)=\int d\omega\,\tilde A(\omega)e^{i\omega t}/(2\pi)$. 
Under this convention, Eq. \eqref{2.5} becomes
\begin{equation}
\left[
m(\omega_p^2-\gamma\omega^2)
+
i\gamma\tilde{\nu}(\omega)\omega
\right]\tilde{s}_{p}(\omega)
=
\Xi_{p,xy}\tilde{\varepsilon}_{xy}(\omega).
\label{2.6}
\end{equation}
The nonaffine contribution to the shear stress follows from the work conjugacy between stress and strain. 
To first order in the strain amplitude,
\begin{equation}
\tilde{\sigma}_{xy}(\omega)
=
-\frac{1}{\mathring V}
\sum_p
\Xi_{p,xy}\tilde{s}_p(\omega),
\label{2.7}
\end{equation}
where $\mathring V$ is the reference volume. 
Combining Eqs. \eqref{2.6} and \eqref{2.7}, and defining $\tilde{\sigma}_{xy}(\omega)=G^*(\omega)\tilde{\varepsilon}_{xy}(\omega)$, gives
\begin{equation}
G^*(\omega)
=
\frac{1}{\mathring V}
\sum_p
\frac{\Xi_{p,xy}^{2}}
{m\gamma\omega^2-m\omega_p^2-i\gamma\tilde{\nu}(\omega)\omega}.
\label{2.8}
\end{equation}
The imaginary part of Eq. \eqref{2.8} is the loss modulus. 
For a genuinely causal bath, the Fourier-transformed memory kernel may be complex. 
We therefore write $\tilde{\nu}(\omega)=\nu_R(\omega)+i\nu_I(\omega)$, where $\nu_R$ is the dissipative part and $\nu_I$ is the reactive part. 
For a passive bath with positive low-frequency dissipation, the dissipative part of the response is
\begin{equation}
G''(\omega)
=
\frac{1}{\mathring V}
\sum_p
\frac{
\Xi_{p,xy}^{2}\gamma\nu_R(\omega)\omega
}{
\left[
m(\omega_p^2-\gamma\omega^2)
-
\gamma\nu_I(\omega)\omega
\right]^2
+
\gamma^2\nu_R^2(\omega)\omega^2
}.
\label{2.9}
\end{equation}
The zero-frequency shear viscosity is defined by $\eta=\lim_{\omega\to0}G''(\omega)/\omega$ \cite{Kubo_1957,Zaccone_2023}. 
In this limit the reactive part does not contribute, and below we write $\tilde{\nu}(0)\equiv\nu_R(0)$. 
Taking the limit in Eq. \eqref{2.9} yields
\begin{equation}
\eta_{\rm unav}(\gamma)
=
\frac{\gamma\tilde{\nu}(0)}{\mathring V}
\sum_p
\frac{\Xi_{p,xy}^{2}}{m^2\omega_p^4}.
\label{2.10}
\end{equation}
No scaling assumption has entered this expression. 
 The approximations at this stage are linear response, the fixed-$\gamma$ scalar transverse reduction described above, the neglect of off-diagonal projected inertia and friction elements, and the restriction to a scalar isotropic memory function in the shear channel.

 Within the stable real-frequency sector just defined, for a macroscopically large disordered system, the discrete normal-mode sum may be converted into an integral over the vibrational or instantaneous-normal-mode density of states. 
We define the shell-averaged affine-force correlator by
\begin{equation}
\Gamma_{xyxy}(\omega)
=
\left\langle
\Xi_{p,xy}^{2}
\right\rangle_{\omega_p\in[\omega,\omega+d\omega]} .
\label{2.11}
\end{equation}
Then Eq. \eqref{2.10} becomes
\begin{equation}
\eta_{\rm unav}(\gamma)
=
\frac{\gamma\tilde{\nu}(0)}{\mathring V}
\int_0^{\omega_D}
\frac{
g(\omega_p)\Gamma_{xyxy}(\omega_p)
}{
m^2\omega_p^4
}
\,d\omega_p .
\label{2.12}
\end{equation}
Equation \eqref{2.12} is the unaveraged relativistic nonaffine viscosity formula. 
Its structure separates three quantities: the mode density $g(\omega_p)$, the affine-force spectral weight $\Gamma_{xyxy}(\omega_p)$, and the zero-frequency memory kernel $\tilde{\nu}(0)$. 
The Lorentz factor multiplies the entire expression only because it has been treated as a fixed kinematic parameter.

\subsection{Debye reduction and the unaveraged Lorentz enhancement}
\label{subsec_debye_reduction_unaveraged_gamma}
To make the dependence on the collective mode spectrum explicit, we now evaluate Eq. \eqref{2.12} under the low-frequency Debye approximation. 
For an isotropic three-dimensional medium with acoustic speed $c_s$,
\begin{equation}
g(\omega_p)
=
\frac{\mathring V\,\omega_p^2}{2\pi^2c_s^3},
\qquad
0\leq \omega_p\leq \omega_D .
\label{2.13}
\end{equation}

The low-frequency form used below is an adapted central-force result from nonaffine normal-mode theory \cite{Lemaitre_2006,Zaccone_2011,Zaccone_2023}. 
For a smooth central pair interaction $U(r)$, the microscopic affine force generated by an $xy$ shear contains the bond factor $[\kappa_{ij}r_{ij}-t_{ij}]n_{ij}^{x}n_{ij}^{y}\hat{\mathbf n}_{ij}$, where $\kappa_{ij}=U''(r_{ij})$ is the local bond curvature and $t_{ij}=U'(r_{ij})$ is the bond tension in the sign convention of the central-force formulation \cite{Zaccone_2023}. 
The reduction used here assumes an instantaneous mechanically stable configuration, an isotropic distribution of bond directions, and a narrow local structural scale that can be represented by one characteristic separation $R_0$ and one effective stiffness $\kappa$. 
We also neglect the explicit prestress contribution or absorb it into the effective stiffness. 
Under these assumptions, the shell-averaged mode covariance has the form $d\kappa R_0^2\lambda_p\sum_\alpha B_{\alpha,xyxy}$, where $d=3$ and $\lambda_p=m\omega_p^2$ \cite{Zaccone_2011,Zaccone_2023}. 
For isotropic bond orientations, $\sum_\alpha B_{\alpha,xyxy}=\langle n_x^2n_y^2\rangle_\Omega=(4\pi)^{-1}\int d\Omega\,n_x^2n_y^2=1/15$. 
The spatial factor $d=3$ therefore gives $3/15=1/5$. 
The shear-channel correlator then becomes

\begin{equation}
\Gamma_{xyxy}(\omega_p)
=
\frac{1}{5}m\kappa R_0^2\omega_p^2 ,
\label{2.14}
\end{equation}

where $R_0$ has dimensions of length and $\kappa$ has dimensions of force per length, equivalently energy per length squared. 
Thus $\kappa R_0^2$ is an energy scale, and the right-hand side of Eq. \eqref{2.14} has dimensions of force squared, as required by the definition in Eq. \eqref{2.11}. 
For a specified smooth interaction and instantaneous structure, $\kappa$ and $R_0$ can in principle be obtained from the pair potential and the local neighbor geometry. 
After the one-scale Debye reduction, they should instead be read as effective local quantities obtained after the configurational and orientational averages that define the shell correlator. 
Equation \eqref{2.14} is therefore a low-frequency central-force closure rather than a universal identity for an arbitrary relativistic gas. 
Its use requires a smooth local interaction scale, isotropy, self-averaging over the frequency shell, and stable acoustic-like modes in the frequency range retained in the Debye approximation. 
An ideal hard-sphere interaction is discontinuous, so $U''(r)$ and a conventional Hessian curvature are not defined at contact. 
Whenever we later use $\kappa R_0^2$ in a hard-sphere scaling argument, we therefore interpret it as an effective collisional or thermal stiffness scale rather than as the literal curvature of the hard-sphere potential.

Here mechanical stability refers only to the time window over which the linear response is evaluated and does not imply permanent rigidity.
The Debye step retains the positive acoustic-like branch of that short-time spectrum.
Negative-Hessian directions and strongly overdamped excitations are not included in Eqs. \eqref{2.13} and \eqref{2.14}.

Substitution of Eqs. \eqref{2.13} and \eqref{2.14} into Eq. \eqref{2.12} gives
\begin{equation}
\eta_{\rm unav}(\gamma)
=
\frac{\gamma\tilde{\nu}(0)}{\mathring V}
\int_0^{\omega_D}
\frac{
\left(\mathring V\omega_p^2/2\pi^2c_s^3\right)
\left(m\kappa R_0^2\omega_p^2/5\right)
}{
m^2\omega_p^4
}
\,d\omega_p .
\label{2.15}
\end{equation}
The powers of $\omega_p$ cancel before the integration is performed. 
Therefore,
\begin{equation}
\eta_{\rm unav}(\gamma)
=
\frac{\gamma\kappa R_0^2\tilde{\nu}(0)}{10\pi^2m c_s^3}
\int_0^{\omega_D}d\omega_p
=
\frac{\gamma\kappa R_0^2\tilde{\nu}(0)\omega_D}{10\pi^2m c_s^3}.
\label{2.16}
\end{equation}
The nonrelativistic limit is obtained by setting $\gamma=1$, while keeping the same nonaffine spectral data:
\begin{equation}
\eta_{\rm NR}
=
\frac{\kappa R_0^2\tilde{\nu}(0)\omega_D}{10\pi^2m c_s^3}.
\label{2.17}
\end{equation}
Equations \eqref{2.16} and \eqref{2.17} imply the fixed-$\gamma$ enhancement law
\begin{equation}
\frac{\eta_{\rm unav}(\gamma)}{\eta_{\rm NR}}
=
\gamma .
\label{2.18}
\end{equation}
This is the cleanest form of the proper-momentum enhancement mechanism. 
It is also the precise point at which the unaveraged theory reaches its limit of validity. 
For a single tagged particle with prescribed speed, Eq. \eqref{2.18} is a meaningful kinematic statement. 
For a thermal gas, however, $\gamma$ is not a state variable but a function of the particle momentum,
\begin{equation}
\gamma(\mathbf p)
=
\frac{p^0}{mc}
=
\sqrt{1+\frac{\mathbf p^2}{m^2c^2}} .
\label{2.19}
\end{equation}
Hence the scalar multiplier in Eq. \eqref{2.12} cannot be the final relativistic thermal answer. 
For any momentum-dependent quantity $A(\mathbf p)$, we use $\langle A\rangle_J$ for the normalized Maxwell-J\"uttner equilibrium average in the local rest frame, defined by $\langle A\rangle_J\equiv\left[\int_{\mathbb R^3}A(\mathbf p)e^{-\zeta\gamma(\mathbf p)}d^3p\right]/\left[\int_{\mathbb R^3}e^{-\zeta\gamma(\mathbf p)}d^3p\right]$ with the standard classical Maxwell-J\"uttner weight \cite{Juttner_1911,Dunkel:2006nk,Dunkel:2008ngc}.
If the memory kernel and the affine-force correlator are independent of momentum, the next step is the replacement $\gamma\mapsto\langle\gamma\rangle_J$. 
If they are momentum dependent, the correct object is instead the thermal average of their product,
\begin{equation}
\gamma\tilde{\nu}(0)\Gamma_{xyxy}(\omega_p)
\longrightarrow
\left\langle
\gamma(\mathbf p)\,
\tilde{\nu}(0,\mathbf p)\,
\Gamma_{xyxy}(\omega_p,\mathbf p)
\right\rangle_J .
\label{2.20}
\end{equation}
Equation \eqref{2.20} is not an additional assumption; it is the minimal consistency requirement imposed by relativistic equilibrium statistics. 
The remainder of the paper develops this replacement explicitly using the Maxwell-J\"uttner measure and then derives the corresponding finite-$\zeta$ viscosity.

\section{Maxwell-J\"uttner Equilibrium and Relativistic Averages}
\label{sec_juttner_average}
\subsection{Choice of relativistic equilibrium measure}
\label{subsec_juttner_measure}
The unaveraged result of Sec. \ref{sec_rel_nonaffine} treats the Lorentz factor as a fixed parameter, whereas a relativistic gas at thermal equilibrium contains particles with a distribution of momenta. 
For a classical one-component gas in the local rest frame of the fluid, the conventional relativistic equilibrium distribution is the Maxwell-J\"uttner distribution \cite{Juttner_1911,Dunkel:2006nk,Dunkel:2008ngc,Ambrus:2023qcl}. 
We use the particle-counting momentum measure $d^3p$, because the shear viscosity is constructed as a particle-averaged response coefficient in the local rest frame. 
With $E_{\mathbf p}=\gamma(\mathbf p)mc^2$, $\gamma(\mathbf p)=\sqrt{1+\mathbf p^2/(m^2c^2)}$, and $\zeta=mc^2/(k_B T)$, the normalized probability density is
\begin{equation}
f_J(\mathbf p,\zeta)
=
\frac{\zeta}{4\pi m^3c^3K_2(\zeta)}
\exp[-\zeta\gamma(\mathbf p)] .
\label{3.1}
\end{equation}
The normalization in Eq. \eqref{3.1} follows by changing variables from $p=|\mathbf p|$ to $\gamma$. 
Since $p=mc\sqrt{\gamma^2-1}$ and $dp=mc\,\gamma(\gamma^2-1)^{-1/2}d\gamma$, the momentum volume element becomes
\begin{equation}
d^3p
=
4\pi m^3c^3\gamma\sqrt{\gamma^2-1}\,d\gamma .
\label{3.2}
\end{equation}
Therefore,
\begin{equation}
\int_{\mathbb R^3}e^{-\zeta\gamma(\mathbf p)}d^3p
=
4\pi m^3c^3
\int_1^\infty
\gamma\sqrt{\gamma^2-1}\,e^{-\zeta\gamma}\,d\gamma
=
4\pi m^3c^3\frac{K_2(\zeta)}{\zeta},
\label{3.3}
\end{equation}
where the last equality is the standard integral representation of the modified Bessel function appearing in relativistic ideal-gas thermodynamics \cite{Juttner_1911,Dunkel:2008ngc,Ambrus:2023qcl}. 
Equation \eqref{3.1} is thus normalized as $\int f_J(\mathbf p,\zeta)d^3p=1$.

This choice should be distinguished from the Lorentz-invariant phase-space measure used in covariant kinetic integrals. 
In relativistic kinetic theory the distribution function is a scalar, but particle-number averages in the local rest frame are obtained from the $N^0$ component of the current, which reduces to a $d^3p$ particle-counting average for a classical gas \cite{Anderson:1974nyl,Rocha:2021zcw,Ambrus:2023qcl}. 
Alternative maximum-entropy constructions based on a Lorentz-invariant reference measure lead to modified J\"uttner distributions with an additional energy-dependent prefactor \cite{Dunkel:2006hc,Dunkel:2006nk}. 
Those alternatives are physically meaningful for specific measure-theoretic questions, but the present work adopts Eq. \eqref{3.1} because it is the standard equilibrium measure used in relativistic kinetic-theory transport calculations.

For any momentum-dependent quantity $A(\mathbf p)$, we define the Maxwell-J\"uttner average by
\begin{equation}
\langle A\rangle_J
=
\int_{\mathbb R^3}A(\mathbf p)f_J(\mathbf p,\zeta)\,d^3p
=
\frac{\zeta}{K_2(\zeta)}
\int_1^\infty
A(\gamma)\gamma\sqrt{\gamma^2-1}\,e^{-\zeta\gamma}\,d\gamma ,
\label{3.4}
\end{equation}
where isotropy has been used in the second equality. 
The reduction to a one-dimensional integral is exact for all functions that depend on momentum only through $\gamma$.

\begin{figure}
\centering
\includegraphics[width=0.60\textwidth]{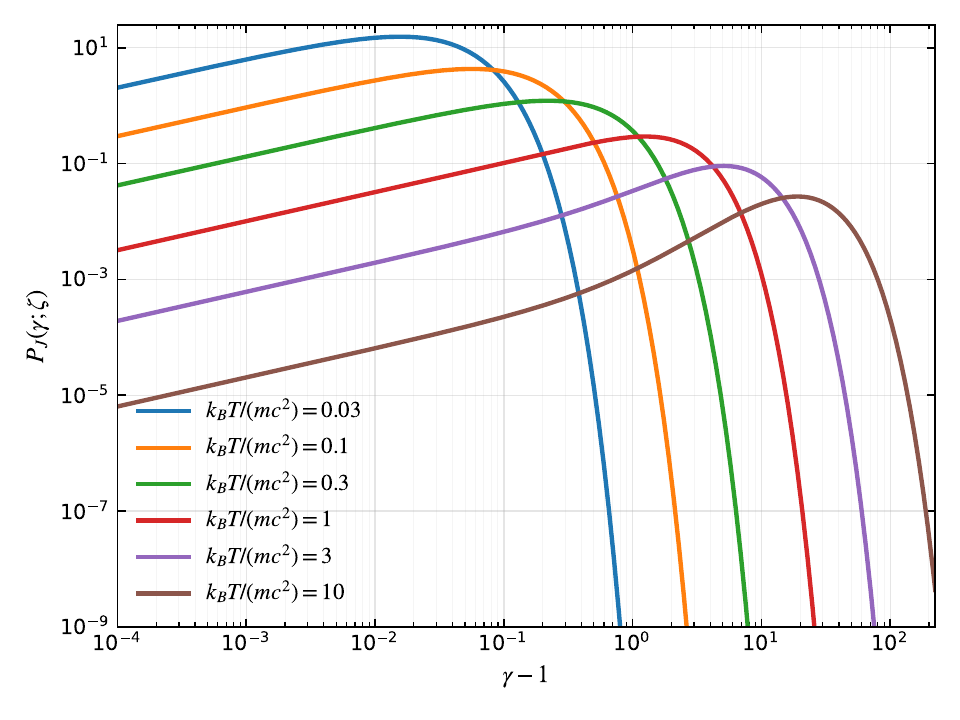}
\caption{
Maxwell-J\"uttner probability density of the Lorentz factor, written as a
density with respect to $d\gamma$, for representative values of the dimensionless
temperature $\theta=k_BT/(mc^2)=1/\zeta$.  The horizontal axis is $\gamma-1$,
the kinetic energy in units of $mc^2$.  At low temperature the distribution is
sharply localized near $\gamma=1$, while in the relativistic regime it broadens
and shifts over several decades in $\gamma-1$.  The figure therefore displays
the statistical origin of the replacement of a fixed Lorentz factor by
Maxwell-J\"uttner averages.
}
\label{fig_mj-gamma-distribution}
\end{figure}
Figure \ref{fig_mj-gamma-distribution} shows explicitly that the thermal
relativistic gas is not characterized by a single Lorentz factor.  In the
nonrelativistic regime, the probability density is concentrated close to
$\gamma=1$, and the fixed-$\gamma$ approximation used in the unaveraged
nonaffine response is recovered as an asymptotic limit.  As $\theta$ approaches
and exceeds unity, the distribution develops substantial support at
$\gamma-1=\mathcal{O}(1)$ and beyond, so that the relevant response cannot be represented
by a prescribed scalar enhancement.  This motivates the use of the moments
$M_n(\zeta)=\langle\gamma^n\rangle_J$ and, in particular, the thermal mean
$M_1(\zeta)$ and variance ${\rm Var}_J(\gamma)$ derived below.

\subsection{Thermal Lorentz moments}
\label{subsec_thermal_lorentz_moments}
The minimal correction to the unaveraged nonaffine theory is obtained by replacing the fixed Lorentz factor by its Maxwell-J\"uttner mean. 
To compute this mean and its higher moments analytically, define
\begin{equation}
I_0(\zeta)
=
\int_1^\infty
\gamma\sqrt{\gamma^2-1}\,e^{-\zeta\gamma}\,d\gamma
=
\frac{K_2(\zeta)}{\zeta}.
\label{3.5}
\end{equation}
Repeated differentiation of $I_0$ generates Lorentz moments because each derivative brings down one power of $-\gamma$. 
Thus
\begin{equation}
M_n(\zeta)
\equiv
\langle \gamma^n\rangle_J
=
\frac{(-1)^n}{I_0(\zeta)}
\frac{d^n I_0(\zeta)}{d\zeta^n}.
\label{3.6}
\end{equation}
For $n=1$, Eq. \eqref{3.6} gives
\begin{equation}
M_1(\zeta)
=
-\frac{d}{d\zeta}\ln\!\left[\frac{K_2(\zeta)}{\zeta}\right]
=
\frac{K_1(\zeta)}{K_2(\zeta)}
+
\frac{3}{\zeta},
\label{3.7}
\end{equation}
where we used $dK_\nu/d\zeta=-K_{\nu-1}-\nu K_\nu/\zeta$. 
Equation \eqref{3.7} is the standard mean-energy relation for a classical relativistic ideal gas, since $\langle E_{\mathbf p}\rangle_J=mc^2M_1(\zeta)$ \cite{Dunkel:2008ngc,Ambrus:2023qcl}.

The second Lorentz moment is needed to estimate the error incurred when a fluctuating Lorentz factor is replaced by its mean. 
From Eq. \eqref{3.6},
\begin{equation}
M_2(\zeta)
=
\frac{1}{I_0(\zeta)}
\frac{d^2 I_0(\zeta)}{d\zeta^2}
=
1
+
\frac{3}{\zeta}\frac{K_1(\zeta)}{K_2(\zeta)}
+
\frac{12}{\zeta^2}.
\label{3.8}
\end{equation}
Consequently, the thermal variance of the Lorentz factor is
\begin{equation}
{\rm Var}_J(\gamma)
=
M_2(\zeta)-M_1^2(\zeta).
\label{3.9}
\end{equation}
This variance controls the first purely kinematic correction to the scalar-$\gamma$ approximation. 
In the nonrelativistic limit, the modified Bessel expansion gives
\begin{equation}
M_1(\zeta)
=
1+\frac{3}{2\zeta}
+\frac{15}{8\zeta^2}
+\mathcal{O}(\zeta^{-3}),
\qquad
{\rm Var}_J(\gamma)
=
\frac{3}{2\zeta^2}
+\mathcal{O}(\zeta^{-3}).
\label{3.10}
\end{equation}
Thus the unaveraged limit $\gamma=1$ is recovered smoothly as $\zeta\to\infty$. 
The first correction $3/(2\zeta)$ is the relativistic rewriting of the nonrelativistic equipartition contribution $\langle p^2/(2m)\rangle=3k_BT/2$.

In the ultrarelativistic limit, the same Bessel representation gives
\begin{equation}
M_1(\zeta)
=
\frac{3}{\zeta}
+\frac{\zeta}{2}
+\mathcal{O}(\zeta^3\ln\zeta),
\qquad
M_2(\zeta)
=
\frac{12}{\zeta^2}
+\mathcal{O}(1).
\label{3.11}
\end{equation}
The leading term $M_1\simeq3/\zeta$ is the classical massless-gas relation $\langle E_{\mathbf p}\rangle_J\simeq3k_BT$. 
This result also explains why the replacement $\gamma\sim k_BT/(mc^2)$ captures only the scaling but not the numerical coefficient of the thermal relativistic enhancement.

\subsection{Averaging prescriptions}
\label{subsec_averaging_prescriptions}
We now specify how the Maxwell-J\"uttner average enters the nonaffine viscosity formula. 
The unaveraged expression in Eq. \eqref{2.12} contains the product $\gamma\tilde{\nu}(0)\Gamma_{xyxy}(\omega_p)$. 
For a thermal gas, the most general local-rest-frame average is therefore
\begin{equation}
\eta_J
=
\frac{1}{\mathring V}
\int_0^{\omega_D}
\frac{g(\omega_p)}{m^2\omega_p^4}
\left\langle
\gamma(\mathbf p)\,
\tilde{\nu}(0,\mathbf p,\omega_p)\,
\Gamma_{xyxy}(\omega_p,\mathbf p)
\right\rangle_J
d\omega_p .
\label{3.12}
\end{equation}
Equation \eqref{3.12} is the formal Maxwell-J\"uttner averaged nonaffine viscosity functional. 
 It is exact only after the scalar transverse mean-field reduction stated in Sec. \ref{sec_rel_nonaffine}. It does not represent the full tensorial thermal many-particle response of Eq. \eqref{2.coupled.modes}. Within that reduction, no assumption has yet been made about the momentum dependence of the memory function or the affine-force correlator.

The simplest closed model is obtained when the memory kernel and the nonaffine force correlator are separable in collective-mode and momentum variables,
\begin{equation}
\tilde{\nu}(0,\mathbf p,\omega_p)\Gamma_{xyxy}(\omega_p,\mathbf p)
=
\tilde{\nu}_0\,\Gamma_0(\omega_p)\,\chi(\gamma),
\label{3.13}
\end{equation}
where $\chi(\gamma)$ is a dimensionless weighting function normalized so that $\chi\to1$ in the nonrelativistic hard-sphere reference limit. 
 Substituting Eq.~\eqref{3.13} into Eq.~\eqref{3.12} gives
\begin{equation}
\eta_J
=
F_\chi(\zeta)
\frac{\tilde{\nu}_0}{\mathring V}
\int_0^{\omega_D}
\frac{g(\omega_p)\Gamma_0(\omega_p)}{m^2\omega_p^4}\,d\omega_p,
\qquad
F_\chi(\zeta)
=
\langle\gamma\chi(\gamma)\rangle_J .
\label{3.14}
\end{equation}
The minimal Maxwell-J\"uttner averaged model corresponds to $\chi(\gamma)=1$, giving
\begin{equation}
F_0(\zeta)
=
M_1(\zeta)
=
\frac{K_1(\zeta)}{K_2(\zeta)}
+
\frac{3}{\zeta}.
\label{3.15}
\end{equation}
This $F_0(\zeta)$ is the kinematic enhancement function that replaces the fixed factor $\gamma$ in Eq. \eqref{2.18}.

The separable model is useful, but it hides correlations between the Lorentz factor and the dynamical response. 
To expose these correlations, define
\begin{equation}
Q(\mathbf p,\omega_p)
=
\tilde{\nu}(0,\mathbf p,\omega_p)\Gamma_{xyxy}(\omega_p,\mathbf p).
\label{3.16}
\end{equation}
Then the averaged product in Eq. \eqref{3.12} decomposes exactly as
\begin{equation}
\left\langle\gamma Q\right\rangle_J
=
\langle\gamma\rangle_J\langle Q\rangle_J
+
{\rm Cov}_J(\gamma,Q),
\qquad
{\rm Cov}_J(\gamma,Q)
=
\left\langle
(\gamma-\langle\gamma\rangle_J)
(Q-\langle Q\rangle_J)
\right\rangle_J .
\label{3.17}
\end{equation}
Equation \eqref{3.17} identifies the condition under which the replacement $\gamma\mapsto M_1(\zeta)$ is controlled: the covariance between relativistic kinematics and the dynamical response weight must be negligible compared with $M_1\langle Q\rangle_J$. 
If high-momentum particles experience systematically shorter or longer memory times, this covariance term is not small and the viscosity cannot be represented by a purely kinematic enhancement function. 
The next section uses the minimal and separable forms as analytically tractable closures, while keeping Eq. \eqref{3.12} as the general reference formula.

\section{Maxwell-J\"uttner Averaged Nonaffine Shear Viscosity}
\label{sec_averaged_viscosity}
\subsection{Formal averaged viscosity functional}
\label{subsec_formal_averaged_functional}
We now construct the shear viscosity by averaging the mode-resolved relativistic nonaffine response over the Maxwell-J\"uttner equilibrium measure introduced in Sec. \ref{sec_juttner_average}. 
For a particle subensemble with fixed momentum $\mathbf p$, the finite-frequency loss modulus follows from the same modal calculation that led to Eq. \eqref{2.9}, except that the Lorentz factor and the dissipative kernel may depend on $\mathbf p$. 
Thus,
\begin{equation}
G''_{\mathbf p}(\omega)
=
\frac{1}{\mathring V}
\int_0^{\omega_D}
\frac{
g(\omega_p)\Gamma_{xyxy}(\omega_p,\mathbf p)
\gamma(\mathbf p)\tilde{\nu}(\omega,\mathbf p,\omega_p)\omega
}{
m^2\left[\omega_p^2-\gamma(\mathbf p)\omega^2\right]^2
+
\gamma^2(\mathbf p)\tilde{\nu}^{\,2}(\omega,\mathbf p,\omega_p)\omega^2
}
\,d\omega_p .
\label{4.1}
\end{equation}
The observable loss modulus in local equilibrium is the Maxwell-J\"uttner average of Eq. \eqref{4.1},
\begin{equation}
G''_{J}(\omega)
=
\int_{\mathbb R^3}
G''_{\mathbf p}(\omega)f_J(\mathbf p,\zeta)\,d^3p .
\label{4.2}
\end{equation}

The interchange of the low-frequency limit with the equilibrium average requires more than pointwise finiteness of $\tilde{\nu}$. 
We assume first that, for almost every $(\mathbf p,\omega_p)$, the dissipative memory function $\tilde{\nu}(\omega,\mathbf p,\omega_p)$ is continuous in $\omega$ in a neighborhood of $\omega=0$ and has a finite limit $\tilde{\nu}(0,\mathbf p,\omega_p)$. 
A sufficient mathematical condition for exchanging the limit with both integrations in Eq. \eqref{4.2} is the existence of some $\omega_*>0$ and a nonnegative function $M(\mathbf p,\omega_p)$, integrable with respect to $f_J(\mathbf p,\zeta)d^3p\,d\omega_p$ on $\mathbb R^3\times[0,\omega_D]$, such that $|G''_{\mathbf p}(\omega)/\omega|\leq M(\mathbf p,\omega_p)$ for $0<|\omega|<\omega_*$. 
The dominated convergence theorem then permits the low-frequency limit to pass through the Maxwell-J\"uttner and mode integrations. 
Physically, this condition excludes a singular spectral contribution at $\omega_p=0$ and any sequence of asymptotically undamped soft modes that destroys uniform integrability as $\omega\to0$. 
For the Debye central-force reduction adopted in Sec. \ref{subsec_debye_reduction_unaveraged_gamma}, $g(\omega_p)$ and $\Gamma_{xyxy}(\omega_p)$ each vanish quadratically at small $\omega_p$, so the limiting factor $g(\omega_p)\Gamma_{xyxy}(\omega_p)/\omega_p^4$ remains finite at the lower endpoint. 
The Maxwell-J\"uttner factor $e^{-\zeta\gamma}$ also makes every polynomial momentum moment finite for $\zeta>0$. 
Hence finite and continuous $\tilde{\nu}$ together with smooth momentum dependence of $\Gamma_{xyxy}$ satisfy the required integrability whenever they admit the majorant stated above. 
If the memory function or affine-force weight develops a nonintegrable momentum singularity, or if singular zero modes are present, the interchange need not hold and the low-frequency limit must instead be evaluated before the equilibrium integration.

Under these assumptions, the shear viscosity is
\begin{equation}
\eta_J
=
\lim_{\omega\to0}\frac{G''_J(\omega)}{\omega}
=
\frac{1}{\mathring V}
\int_0^{\omega_D}
\frac{g(\omega_p)}
{m^2\omega_p^4}
\left\langle
\gamma(\mathbf p)
\tilde{\nu}(0,\mathbf p,\omega_p)
\Gamma_{xyxy}(\omega_p,\mathbf p)
\right\rangle_J
d\omega_p .
\label{4.3}
\end{equation}
Equation \eqref{4.3} is the Maxwell-J\"uttner averaged nonaffine shear viscosity functional. 
It is the averaged counterpart of Eq. \eqref{2.12}. 
The average is not applied only to the Lorentz factor unless the remaining dynamical quantities are independent of momentum. 
This distinction is essential: the Lorentz factor is kinematic, while $\tilde{\nu}$ and $\Gamma_{xyxy}$ encode microscopic dissipation and affine-force spectral weights.

It is useful to isolate the collective-mode measure from the thermal momentum average. 
Define the mode weight
\begin{equation}
\mathcal A(\omega_p)
=
\frac{g(\omega_p)}{\mathring V m^2\omega_p^4},
\label{4.4}
\end{equation}
and the momentum-dependent dynamical weight
\begin{equation}
Q(\mathbf p,\omega_p)
=
\tilde{\nu}(0,\mathbf p,\omega_p)
\Gamma_{xyxy}(\omega_p,\mathbf p).
\label{4.5}
\end{equation}
Then Eq. \eqref{4.3} becomes
\begin{equation}
\eta_J
=
\int_0^{\omega_D}
\mathcal A(\omega_p)
\left\langle
\gamma(\mathbf p)Q(\mathbf p,\omega_p)
\right\rangle_J
d\omega_p .
\label{4.6}
\end{equation}
This compact form will be the reference expression for all closures below. 
It shows that the averaged viscosity is controlled by a mixed equilibrium moment, not by a universal replacement $\gamma\mapsto\langle\gamma\rangle_J$.

\subsection{Separable closure and effective relativistic enhancement}
\label{subsec_separable_closure}
A closed analytic expression is obtained when the momentum dependence of the dynamical weight is separable from its collective-mode dependence. 
We assume
\begin{equation}
Q(\mathbf p,\omega_p)
=
Q_0(\omega_p)\chi(\gamma),
\qquad
Q_0(\omega_p)
=
\tilde{\nu}_0\,\Gamma_0(\omega_p),
\label{4.7}
\end{equation}
where $\chi(\gamma)$ is dimensionless and normalized by $\chi(\gamma)\to1$ in the nonrelativistic reference limit. 
Substituting Eq. \eqref{4.7} into Eq. \eqref{4.6} gives
\begin{equation}
\eta_J
=
F_\chi(\zeta)
\int_0^{\omega_D}
\mathcal A(\omega_p)Q_0(\omega_p)\,d\omega_p,
\qquad
F_\chi(\zeta)
=
\langle\gamma\chi(\gamma)\rangle_J .
\label{4.8}
\end{equation}
The function $F_\chi(\zeta)$ is the relativistic enhancement factor associated with the chosen dynamical closure. 
It contains the purely kinematic enhancement if $\chi=1$, and it contains additional microscopic weighting if the memory kernel or the affine-force correlator is momentum selective.

The minimal closure is obtained by taking $\chi(\gamma)=1$. 
Using Eq. \eqref{3.7}, the enhancement factor is then
\begin{equation}
F_0(\zeta)
=
\langle\gamma\rangle_J
=
\frac{K_1(\zeta)}{K_2(\zeta)}
+
\frac{3}{\zeta}.
\label{4.9}
\end{equation}
Therefore the minimally averaged nonaffine viscosity is
\begin{equation}
\eta_J^{(0)}
=
\left[
\frac{K_1(\zeta)}{K_2(\zeta)}
+
\frac{3}{\zeta}
\right]
\eta_{\rm NA}^{(0)},
\qquad
\eta_{\rm NA}^{(0)}
=
\int_0^{\omega_D}
\mathcal A(\omega_p)Q_0(\omega_p)\,d\omega_p .
\label{4.10}
\end{equation}
Here $\eta_{\rm NA}^{(0)}$ is the nonaffine viscosity computed from the same collective spectrum and memory amplitude but without relativistic thermal enhancement. 
 Equation \eqref{4.10} is the first central result within the scalar transverse mean-field branch. It replaces the fixed Lorentz factor of that branch by its exact Maxwell-J\"uttner average $F_0(\zeta)$. The statement of exactness applies to this thermal average and not to the tensorial many-particle reduction discussed in Sec. \ref{sec_rel_nonaffine}.

To make contact with the Debye reduction of Sec. \ref{subsec_debye_reduction_unaveraged_gamma}, take the acoustic density of states and the shear affine-force correlator,
\begin{equation}
g(\omega_p)
=
\frac{\mathring V\omega_p^2}{2\pi^2c_s^3},
\qquad
\Gamma_0(\omega_p)
=
\frac{1}{5}m\kappa R_0^2\omega_p^2 .
\label{4.11}
\end{equation}
With $\tilde{\nu}_0$ independent of $\omega_p$, Eq. \eqref{4.10} gives
\begin{equation}
\eta_J^{(0)}
=
F_0(\zeta)
\frac{\tilde{\nu}_0}{\mathring V}
\int_0^{\omega_D}
\frac{
\left(\mathring V\omega_p^2/2\pi^2c_s^3\right)
\left(m\kappa R_0^2\omega_p^2/5\right)
}{
m^2\omega_p^4
}
\,d\omega_p .
\label{4.12}
\end{equation}
The $\omega_p$ powers cancel exactly, so that
\begin{equation}
\eta_J^{(0)}
=
F_0(\zeta)
\frac{\kappa R_0^2\tilde{\nu}_0\omega_D}{10\pi^2m c_s^3}.
\label{4.13}
\end{equation}
Comparing Eq. \eqref{4.13} with Eq. \eqref{2.17}, we obtain
\begin{equation}
\frac{\eta_J^{(0)}}{\eta_{\rm NR}}
=
F_0(\zeta)
=
\frac{K_1(\zeta)}{K_2(\zeta)}
+
\frac{3}{\zeta}.
\label{4.14}
\end{equation}
Equation \eqref{4.14} is the averaged replacement of the unaveraged enhancement law $\eta_{\rm unav}/\eta_{\rm NR}=\gamma$. 
It reduces to unity in the nonrelativistic limit and grows proportionally to $1/\zeta$ in the ultrarelativistic limit, with the precise coefficients determined by the Maxwell-J\"uttner distribution.

\subsection{Nonseparable correction hierarchy}
\label{subsec_nonseparable_corrections}
The minimal result above is exact only if the dynamical weight is independent of momentum, or if its momentum dependence factorizes into a prescribed $\chi(\gamma)$. 
For a general memory kernel and a general affine-force correlator, the correction to the minimal enhancement can be written without approximation. 
Using Eq. \eqref{3.17} in Eq. \eqref{4.6}, we obtain
\begin{equation}
\eta_J
=
M_1(\zeta)
\int_0^{\omega_D}
\mathcal A(\omega_p)
\langle Q(\mathbf p,\omega_p)\rangle_J
d\omega_p
+
\int_0^{\omega_D}
\mathcal A(\omega_p)
{\rm Cov}_J[\gamma,Q(\omega_p)]
d\omega_p .
\label{4.15}
\end{equation}
The first term is the mean-field Maxwell-J\"uttner enhancement, while the second term measures the correlation between relativistic kinematics and microscopic dissipation. 
If faster particles have a larger effective memory weight, the covariance term is positive; if they decorrelate more rapidly, it is negative.

A systematic hierarchy follows by expanding the normalized dynamical weight around its Maxwell-J\"uttner mean at fixed $\omega_p$. 
Let
\begin{equation}
Q(\mathbf p,\omega_p)
=
\bar Q(\omega_p)
\left[
1
+
a_1(\omega_p)\delta\gamma
+
\frac{a_2(\omega_p)}{2}
\left(\delta\gamma^2-\mu_2\right)
+
\mathcal{O}(\delta\gamma^3)
\right],
\label{4.16}
\end{equation}
where $\bar Q(\omega_p)=\langle Q(\mathbf p,\omega_p)\rangle_J$, $\delta\gamma=\gamma-M_1$, and $\mu_n=\langle(\delta\gamma)^n\rangle_J$. 
The subtraction of $\mu_2$ ensures that $\langle Q\rangle_J=\bar Q+\mathcal{O}(\delta\gamma^3)$. 
Multiplying Eq. \eqref{4.16} by $\gamma=M_1+\delta\gamma$ and averaging gives
\begin{equation}
\left\langle\gamma Q(\omega_p)\right\rangle_J
=
\bar Q(\omega_p)
\left[
M_1
+
a_1(\omega_p)\mu_2
+
\frac{a_2(\omega_p)}{2}\mu_3
+
\mathcal{O}(\mu_4)
\right].
\label{4.17}
\end{equation}
Substitution into Eq. \eqref{4.6} yields the nonseparable correction expansion
\begin{equation}
\eta_J
=
\int_0^{\omega_D}
\mathcal A(\omega_p)\bar Q(\omega_p)
\left[
M_1
+
a_1(\omega_p)\mu_2
+
\frac{a_2(\omega_p)}{2}\mu_3
+
\mathcal{O}(\mu_4)
\right]
d\omega_p .
\label{4.18}
\end{equation}

Equation \eqref{4.18} organizes departures from the factorized Maxwell-J\"uttner average, but a finite truncation is not by itself an accuracy statement. 
The central moments are small when $\zeta\gg1$, while they grow in the relativistic regime. 
Consequently, the coefficients $a_1,a_2,\ldots$ cannot be assigned representative values and then interpreted as a quantitative prediction. 
A specific momentum-dependent response is needed before the size of the covariance can be assessed. 
We therefore remove the purely illustrative coefficient curves and retain Eq. \eqref{4.18} only as a local expansion. 
A quantitative assessment therefore requires a specified momentum-dependent shear weight rather than representative covariance coefficients.

Equations \eqref{4.3}, \eqref{4.14}, and \eqref{4.18} summarize the three levels of the theory. 
Equation \eqref{4.3} is the formal averaged viscosity functional. 
Equation \eqref{4.14} is the minimal closed enhancement law. 
Equation \eqref{4.18} is the local correction hierarchy that organizes the failure of factorization. 
The limiting behavior of these expressions is analyzed next.

\section{Limiting Regimes and Scaling Laws}
\label{sec_limits}
\subsection{Nonrelativistic limit}
\label{subsec_nonrelativistic_limit}
The minimally averaged viscosity obtained in Eq. \eqref{4.13} may be written as
\begin{equation}
\eta_J^{(0)}(T)
=
F_0(\zeta)\,
\eta_{\rm NA}^{(0)}(T),
\qquad
F_0(\zeta)
=
\frac{K_1(\zeta)}{K_2(\zeta)}
+
\frac{3}{\zeta},
\qquad
\zeta=\frac{mc^2}{k_B T}.
\label{5.1}
\end{equation}
Here $\eta_{\rm NA}^{(0)}$ contains the collective nonaffine spectrum, the zero-frequency memory kernel, and the acoustic parameters, whereas $F_0$ contains only the relativistic thermal enhancement. 
In the Debye closure of Sec. \ref{subsec_separable_closure},
\begin{equation}
\eta_{\rm NA}^{(0)}(T)
=
\frac{\kappa(T)R_0^2(T)\tilde{\nu}_0(T)\omega_D(T)}
{10\pi^2m c_s^3(T)} .
\label{5.2}
\end{equation}
The separation in Eq. \eqref{5.1} is useful because the nonrelativistic limit is controlled by $F_0(\zeta)$ alone, independently of the chosen model for $\eta_{\rm NA}^{(0)}$.

For $\zeta\gg1$, the modified Bessel functions have the asymptotic expansion used in relativistic equilibrium thermodynamics \cite{Juttner_1911,Dunkel:2008ngc,Ambrus:2023qcl}. 
Using Eq. \eqref{3.10}, Eq. \eqref{5.1} gives
\begin{equation}
F_0(\zeta)
=
1
+
\frac{3}{2\zeta}
+
\frac{15}{8\zeta^2}
+
\mathcal{O}(\zeta^{-3}).
\label{5.3}
\end{equation}
Therefore,
\begin{equation}
\eta_J^{(0)}
=
\eta_{\rm NA}^{(0)}
\left[
1
+
\frac{3k_B T}{2mc^2}
+
\frac{15}{8}
\left(\frac{k_B T}{mc^2}\right)^2
+
O\!\left(\frac{k_B T}{mc^2}\right)^3
\right].
\label{5.4}
\end{equation}
The first term is the nonrelativistic nonaffine viscosity. 
The leading correction is of order $k_BT/(mc^2)$ and is therefore negligible when the thermal kinetic energy is small compared with the rest energy. 
This proves explicitly that the Maxwell-J\"uttner averaged theory recovers the  nonrelativistic nonaffine response as $\zeta\to\infty$.

 As a scaling consistency test with the dilute hard-sphere law discussed in the relativistic nonaffine construction \cite{Zaccone_2024}, we impose the effective temperature dependences
\begin{equation}
\kappa R_0^2\simeq C_\kappa k_B T,
\qquad
\omega_D\simeq C_D\frac{k_B T}{\hbar},
\qquad
c_s\simeq C_s\sqrt{\frac{k_B T}{m}},
\qquad
\tilde{\nu}_0\simeq C_\nu\frac{\hbar}{d^2},
\label{5.5}
\end{equation}

where $d$ is the hard-sphere diameter. 
In this hard-sphere use, $C_\kappa k_B T$ defines the effective collisional stiffness scale introduced below Eq. \eqref{2.14}. 
It does not represent $U''(R_0)R_0^2$, because an ideal hard-sphere interaction has no smooth curvature at contact. 
The constants $C_\kappa,C_D,C_s,C_\nu$ encode the chosen effective stiffness scale, cutoff frequency, acoustic speed, and zero-frequency memory amplitude.

Substitution of Eq. \eqref{5.5} into Eq. \eqref{5.2} gives
\begin{equation}
\eta_{\rm NA}^{(0)}
\simeq
\frac{C_\kappa C_D C_\nu}{10\pi^2 C_s^3}
\frac{\sqrt{mk_B T}}{d^2}.
\label{5.6}
\end{equation}
 These assumed temperature dependences reproduce the nonrelativistic square-root hard-sphere law before the relativistic correction is applied. This is a consistency test of the chosen effective scalings. It is not an independent microscopic derivation from hard-sphere normal modes or from the discontinuous hard-sphere interaction. 
Combining Eqs. \eqref{5.4} and \eqref{5.6}, the averaged relativistic correction to the hard-sphere scaling is
\begin{equation}
\eta_J^{(0)}
\simeq
\frac{C_\kappa C_D C_\nu}{10\pi^2 C_s^3}
\frac{\sqrt{mk_B T}}{d^2}
\left[
1
+
\frac{3k_B T}{2mc^2}
+
O\!\left(
\left[
\frac{k_B T}{mc^2}
\right]^2
\right)
\right].
\label{5.7}
\end{equation}
The constants in Eq. \eqref{5.7} are not fixed by the scaling theory.  Their values, and any assumed weak temperature dependence, must be supplied independently for a specified hard-sphere realization. 
They require either a microscopic calculation of the memory kernel or calibration against a kinetic-theory or molecular-dynamics reference system. 
The important point is that the relativistic averaging modifies the classical result multiplicatively and analytically in powers of $k_BT/(mc^2)$.

\subsection{Moderately relativistic crossover}
\label{subsec_moderately_relativistic_crossover}
The regime $\zeta=\mathcal{O}(1)$ is the genuinely new domain of the present construction. 
In this regime neither $\gamma\simeq1$ nor $\gamma\simeq k_BT/(mc^2)$ is accurate, and the full Bessel-function expression in Eq. \eqref{5.1} must be retained. 
A useful check is the logarithmic temperature slope of the enhancement factor,
\begin{equation}
\mathcal S_F(\zeta)
\equiv
\frac{d\ln F_0}{d\ln T}.
\label{5.8}
\end{equation}
Since $T\,d/dT=-\zeta\,d/d\zeta$ and $F_0=\langle\gamma\rangle_J$, differentiation of the Maxwell-J\"uttner average gives
\begin{equation}
\frac{dF_0}{d\zeta}
=
-\left[
\langle\gamma^2\rangle_J
-
\langle\gamma\rangle_J^2
\right]
=
-{\rm Var}_J(\gamma).
\label{5.9}
\end{equation}
Therefore,
\begin{equation}
\mathcal S_F(\zeta)
=
\frac{\zeta\,{\rm Var}_J(\gamma)}{F_0(\zeta)}.
\label{5.10}
\end{equation}
Equation \eqref{5.10} is exact for the minimal Maxwell-J\"uttner closure. 
It shows that the relativistic enhancement contributes a temperature-dependent effective power governed by the thermal width of the Lorentz-factor distribution.

The total logarithmic slope of the viscosity is obtained by adding the slope of the nonaffine background:
\begin{equation}
\frac{d\ln\eta_J^{(0)}}{d\ln T}
=
\frac{d\ln\eta_{\rm NA}^{(0)}}{d\ln T}
+
\frac{\zeta\,{\rm Var}_J(\gamma)}{F_0(\zeta)} .
\label{5.11}
\end{equation}
If the nonaffine background is locally approximated by a power law $\eta_{\rm NA}^{(0)}\propto T^{\alpha_{\rm NA}}$, then
\begin{equation}
\eta_J^{(0)}
\ \hbox{has local exponent}\ 
\alpha_{\rm eff}(\zeta)
=
\alpha_{\rm NA}
+
\frac{\zeta\,{\rm Var}_J(\gamma)}{F_0(\zeta)} .
\label{5.12}
\end{equation}
This relation is more informative than assigning a single universal power of $T$. 
It shows that the relativistic correction evolves continuously from a negligible contribution in the nonrelativistic domain to an additional power of $T$ in the ultrarelativistic domain.

Using Eqs. \eqref{3.10} and \eqref{3.11}, the slope contribution has the limiting values
\begin{equation}
\frac{\zeta\,{\rm Var}_J(\gamma)}{F_0(\zeta)}
=
\frac{3}{2\zeta}
+
\mathcal{O}(\zeta^{-2}),
\qquad
\zeta\gg1,
\label{5.13}
\end{equation}
and
\begin{equation}
\frac{\zeta\,{\rm Var}_J(\gamma)}{F_0(\zeta)}
=
1
+
\mathcal{O}(\zeta^2\ln\zeta),
\qquad
\zeta\ll1.
\label{5.14}
\end{equation}
Thus the averaged Lorentz enhancement does not abruptly change the scaling law; it supplies a smooth crossover in the effective temperature exponent. 
This is the main advantage of the Maxwell-J\"uttner averaged formulation over the unaveraged estimate $\gamma\sim k_BT/(mc^2)$.

\begin{figure}
    \centering
    \includegraphics[width=0.48\textwidth]{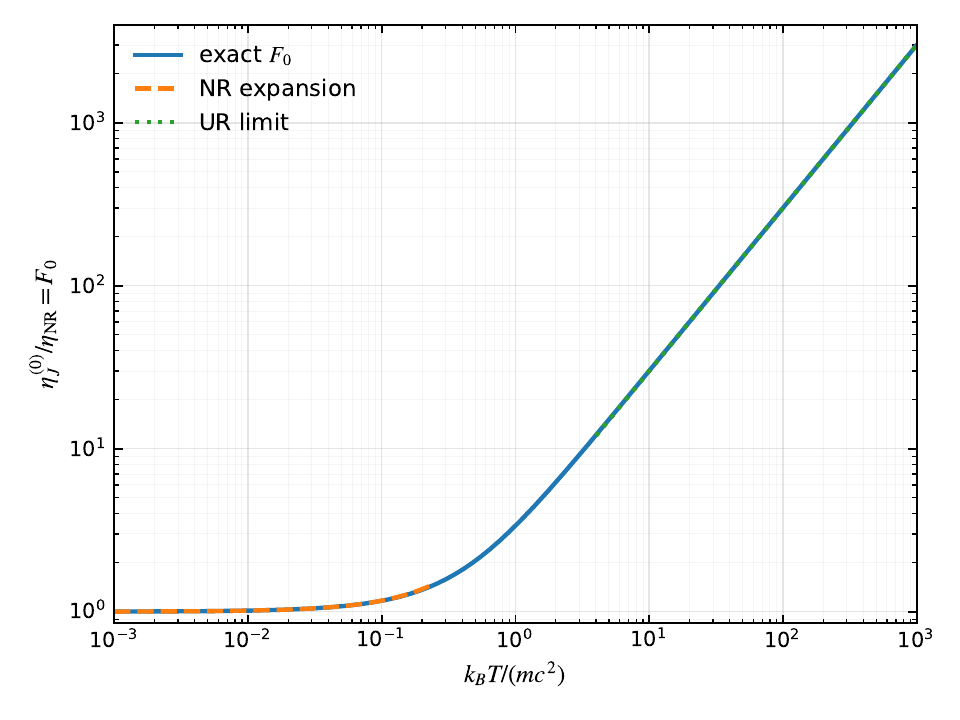}
    \includegraphics[width=0.48\textwidth]{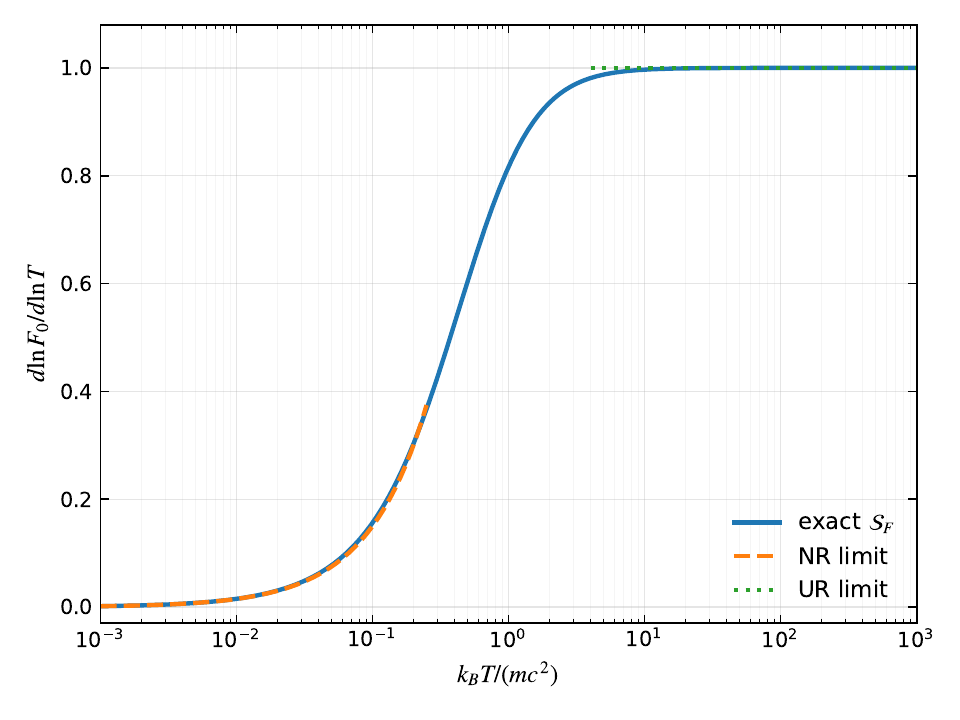}
    \caption{Maxwell-J\"uttner enhancement of the minimally averaged nonaffine shear viscosity.
The left panel shows the exact ratio $\eta_J^{(0)}/\eta_{\rm NR}=F_0(\zeta)$, with
$F_0(\zeta)=K_1(\zeta)/K_2(\zeta)+3/\zeta$ and $\zeta=mc^2/(k_BT)$, compared with its
nonrelativistic and ultrarelativistic limiting forms. The right panel shows the logarithmic
temperature slope $\mathcal S_F=d\ln F_0/d\ln T=\zeta\,{\rm Var}_J(\gamma)/F_0$.
The crossover is smooth: the relativistic correction is negligible for $k_BT\ll mc^2$,
while it contributes one additional power of temperature in the ultrarelativistic regime.
 Both panels follow directly from exact Maxwell-J\"uttner moments and contain no illustrative fit parameters.}
    \label{fig_mj-enhancement}
\end{figure}
Figure \ref{fig_mj-enhancement} displays the exact Maxwell-J\"uttner enhancement factor and its logarithmic temperature slope. The left panel shows that the minimally averaged result interpolates continuously between the nonrelativistic value $F_0\to1$ and the ultrarelativistic growth $F_0\simeq3k_BT/(mc^2)$. The right panel gives the corresponding local exponent contribution $\mathcal S_F=d\ln F_0/d\ln T$, which is controlled by the thermal variance of the Lorentz factor. This representation makes clear that the Maxwell-J\"uttner averaging does not impose a sharp transition between nonrelativistic and ultrarelativistic regimes. Instead, it supplies a smooth finite-mass crossover, so that the effective scaling of $\eta_J^{(0)}$ is the sum of the nonaffine background exponent and the kinematic contribution $\mathcal S_F$.

The same analysis extends to a separable dynamical weighting $\chi(\gamma)$. 
For $F_\chi=\langle\gamma\chi(\gamma)\rangle_J$, differentiation gives
\begin{equation}
\frac{d\ln F_\chi}{d\ln T}
=
\frac{
\zeta\,{\rm Cov}_J\!\left(\gamma,\gamma\chi(\gamma)\right)
}{
F_\chi(\zeta)
},
\label{5.15}
\end{equation}
provided $\chi$ has no explicit temperature dependence other than through $\gamma$. 
Equation \eqref{5.15} shows how a momentum-selective memory kernel changes the crossover. 
If $\chi$ weights high-$\gamma$ particles more strongly, the crossover steepens; if it suppresses them, the crossover weakens.

\subsection{Ultrarelativistic limit}
\label{subsec_ultrarelativistic_limit}
In the ultrarelativistic limit $\zeta\ll1$, the leading Maxwell-J\"uttner enhancement is
\begin{equation}
F_0(\zeta)
=
\frac{3}{\zeta}
+
\mathcal{O}(\zeta)
=
\frac{3k_B T}{mc^2}
+
O\!\left(\frac{mc^2}{k_B T}\right).
\label{5.16}
\end{equation}
Substitution into the Debye-reduced viscosity formula gives
\begin{equation}
\eta_J^{(0)}
=
\frac{3k_B T}{mc^2}
\frac{\kappa R_0^2\tilde{\nu}_0\omega_D}
{10\pi^2m c_s^3}
\left[
1
+
\mathcal{O}(\zeta^2)
\right].
\label{5.17}
\end{equation}
Equation \eqref{5.17} displays the origin of the extra ultrarelativistic power of temperature: it is the Maxwell-J\"uttner average of the Lorentz factor, not an independently imposed hydrodynamic assumption.

To extract a temperature law, the remaining microscopic scales must be specified. 
For a dense ultrarelativistic classical fluid with a temperature-independent sound speed and a thermal microscopic cutoff, take
\begin{equation}
\kappa R_0^2\simeq A_\kappa k_B T,
\qquad
\omega_D\simeq A_D\frac{k_B T}{\hbar},
\qquad
c_s\simeq A_s c,
\qquad
\tilde{\nu}_0\simeq \tilde{\nu}_\infty(T).
\label{5.18}
\end{equation}
Substituting Eq. \eqref{5.18} into Eq. \eqref{5.17} yields
\begin{equation}
\eta_J^{(0)}
\simeq
\frac{3A_\kappa A_D}{10\pi^2A_s^3}
\frac{\tilde{\nu}_\infty(T)}{\hbar}
\frac{(k_BT)^3}{m^2c^5}.
\label{5.19}
\end{equation}
Thus the cubic law follows if $\tilde{\nu}_\infty(T)/\hbar$ is approximately temperature independent in the high-temperature interval under consideration. 
More generally, if $\tilde{\nu}_\infty(T)\propto T^\beta$, then
\begin{equation}
\eta_J^{(0)}
\propto
T^{3+\beta}.
\label{5.20}
\end{equation}
This is the precise sense in which the cubic law is conditional. 
The Maxwell-J\"uttner average supplies one factor of $T$, while the thermal stiffness and cutoff in Eq. \eqref{5.18} supply two further powers. 
Any additional temperature dependence of the memory kernel changes the final exponent.

The result may be expressed as a ratio to entropy density when a classical ultrarelativistic equation of state is specified. 
If $s(T)=C_{\rm eos}k_B(k_BT)^3/(\hbar^3c^3)$ at fixed degeneracy and vanishing chemical potential, then Eq. \eqref{5.19} gives
\begin{equation}
\frac{\eta_J^{(0)}}{s}
\propto
\frac{\hbar^2\tilde{\nu}_\infty(T)}
{k_Bm^2c^2},
\label{5.21}
\end{equation}
up to the dimensionless equation-of-state and cutoff normalizations. 
The ratio is therefore not predicted universally by the kinematic averaging alone. 
It is controlled by the dynamical memory amplitude and by the effective mass scale entering the classical particle description. 
This limitation is important for applications to QGP-like matter, where quantum statistics, screening, pair production, and gauge interactions can modify the memory kernel and the relevant quasiparticle mass scale \cite{Denicol:2012cn,Rocha:2023hts,Ambrus:2023qcl}.

Combining the three regimes, the minimal averaged theory yields the controlled interpolation
\begin{equation}
\eta_J^{(0)}(T)
=
\eta_{\rm NA}^{(0)}(T)
\left[
\frac{K_1(mc^2/k_BT)}{K_2(mc^2/k_BT)}
+
\frac{3k_BT}{mc^2}
\right].
\label{5.22}
\end{equation}
Equation \eqref{5.22} is the finite-mass crossover formula. 
It reduces to the nonrelativistic nonaffine viscosity with analytic relativistic corrections for $k_BT\ll mc^2$, and to an additional linear-in-$T$ enhancement of the nonaffine background for $k_BT\gg mc^2$. 
The remaining temperature dependence is not kinematic; it is determined by the microscopic closure for $\kappa R_0^2$, $\omega_D$, $c_s$, and $\tilde{\nu}_0$.

\section{Comparison with Relativistic Kinetic-Theory Examples}
\label{sec_kinetic_examples}
\subsection{Anderson-Witting and improved relaxation-time examples}
\label{subsec_aw_rta_examples}
We first compare the Maxwell-J\"uttner averaged nonaffine formula with the simplest relativistic kinetic-theory structure: a single-relaxation-time approximation to the Boltzmann equation. 
In this subsection only, we set $c=\hbar=k_B=1$ to avoid obscuring the kinetic-theory weights. 
The relativistic Boltzmann equation for a classical one-component gas is written as
\begin{equation}
p^\mu\partial_\mu f(x,p)=C[f],
\label{6.1}
\end{equation}
where $f$ is the one-particle distribution and $C[f]$ is the collision functional \cite{Anderson:1974nyl,Rocha:2021zcw,Ambrus:2023qcl}. 
The Anderson-Witting relaxation-time model replaces the linearized collision term by
\begin{equation}
C_{\rm AW}[f]
=
-\frac{u_\mu p^\mu}{\tau_R}
\left(f-f_0\right),
\label{6.2}
\end{equation}
where $u^\mu$ is the local fluid four-velocity, $f_0=\exp[-\beta u_\mu p^\mu+\alpha]$ is the local Maxwell-J\"uttner distribution, and $\tau_R$ is a relaxation time \cite{Anderson:1974nyl}. 
The improved relaxation-time construction of Rocha, Denicol, and Noronha modifies the projection structure of the relaxation operator so that microscopic and macroscopic conservation laws are respected for more general matching conditions \cite{Rocha:2021zcw}. 
For the present comparison, the important point is not the detailed matching prescription, but the fact that kinetic relaxation acts on irreducible momentum-space tensors rather than on the scalar Lorentz factor alone.

To see the shear weight explicitly, linearize Eq. \eqref{6.1} around $f_0$ by writing $f=f_0+\delta f$. 
At first order in gradients, and retaining only the shear tensor $\sigma_{\mu\nu}=\nabla_{\langle\mu}u_{\nu\rangle}$, the streaming term is
\begin{equation}
p^\mu\partial_\mu f_0
=
-\beta f_0\,p^{\langle\mu}p^{\nu\rangle}\sigma_{\mu\nu}
+\hbox{terms in the scalar and vector channels}.
\label{6.3}
\end{equation}
Substitution into Eq. \eqref{6.2} gives the shear part of the nonequilibrium correction,
\begin{equation}
\delta f_{\rm sh}
=
\frac{\tau_R\beta}{u_\lambda p^\lambda}
f_0\,p^{\langle\mu}p^{\nu\rangle}\sigma_{\mu\nu}.
\label{6.4}
\end{equation}
The shear-stress tensor is
\begin{equation}
\pi^{\mu\nu}
=
\int dP\,p^{\langle\mu}p^{\nu\rangle}\delta f,
\qquad
dP=
\frac{g\,d^3p}{(2\pi)^3p^0},
\label{6.5}
\end{equation}
where $g$ is the degeneracy factor. 
Using isotropy in the local rest frame, Eq. \eqref{6.4} yields
\begin{equation}
\eta_{\rm RTA}
=
\frac{\tau_R}{15T}
\int
\frac{g\,d^3p}{(2\pi)^3}
\frac{|\mathbf p|^4}{E_{\mathbf p}^2}
f_0(\mathbf p).
\label{6.6}
\end{equation}
Equation \eqref{6.6} is the kinetic-theory analogue of our averaged nonaffine expression. 
It differs structurally from Eq. \eqref{4.3}: the kinetic shear channel weights a particle by $|\mathbf p|^4/E_{\mathbf p}^2$, whereas the minimal nonaffine averaging weights the response by $\gamma=E_{\mathbf p}/m$. 
This difference is not a contradiction. 
It reflects that Eq. \eqref{6.6} is a collision-operator result for the shear tensor, while Eq. \eqref{4.3} is a mode-resolved nonaffine response formula with the dynamical information placed in $\tilde{\nu}\Gamma_{xyxy}$.

The comparison becomes transparent by writing Eq. \eqref{6.6} as a Maxwell-J\"uttner average. 
Let $n=\int g\,d^3p\,f_0/(2\pi)^3$ be the particle density and $E_{\mathbf p}=m\gamma$. 
Then
\begin{equation}
\eta_{\rm RTA}
=
\frac{n\tau_R m^2}{15T}
\left\langle
\frac{(\gamma^2-1)^2}{\gamma^2}
\right\rangle_J
=
n\tau_R m\,
\mathcal R_\eta(\zeta),
\label{6.7}
\end{equation}
with
\begin{equation}
\mathcal R_\eta(\zeta)
=
\frac{\zeta}{15}
\left\langle
\frac{(\gamma^2-1)^2}{\gamma^2}
\right\rangle_J .
\label{6.8}
\end{equation}
In the nonrelativistic limit, $\gamma=1+|\mathbf p|^2/(2m^2)+\mathcal{O}(|\mathbf p|^4/m^4)$ and the Maxwellian moment $\langle |\mathbf p|^4\rangle=15m^2T^2$ gives
\begin{equation}
\mathcal R_\eta(\zeta)
=
\frac{1}{\zeta}
+
\mathcal{O}(\zeta^{-2}),
\qquad
\eta_{\rm RTA}
=
nT\tau_R
+
\mathcal{O}(T^2/m).
\label{6.9}
\end{equation}
In the ultrarelativistic limit, $(\gamma^2-1)^2/\gamma^2=\gamma^2+\mathcal{O}(1)$ and $\langle\gamma^2\rangle_J=12/\zeta^2+\mathcal{O}(1)$, so
\begin{equation}
\mathcal R_\eta(\zeta)
=
\frac{4}{5\zeta}
+
\mathcal{O}(\zeta),
\qquad
\eta_{\rm RTA}
=
\frac{4}{5}P\tau_R
+
\mathcal{O}(m^2/T^2),
\label{6.10}
\end{equation}
where $P=nT$ for a classical ideal gas in the ultrarelativistic limit. 
Equation \eqref{6.10} is the familiar Anderson-Witting shear-viscosity relation \cite{Anderson:1974nyl,Rocha:2021zcw}.

\subsection{Quantitative factorization test from the Anderson-Witting shear weight}
\label{subsec_aw_factorization_test}
The shear weight in Eq. \eqref{6.7} supplies a concrete momentum-dependent response with which we can test the factorization in Eq. \eqref{4.15}. 
We define
\begin{equation}
Q_{\rm AW}(\gamma)
=
\frac{(\gamma^2-1)^2}{\gamma^3},
\qquad
\gamma Q_{\rm AW}(\gamma)
=
\frac{(\gamma^2-1)^2}{\gamma^2}.
\label{6.awq}
\end{equation}
The product on the right is exactly the one-particle shear weight entering Eq. \eqref{6.7}. 
The exact factorization ratio is therefore
\begin{equation}
\mathcal F_{\rm AW}(\zeta)
=
\frac{\langle\gamma Q_{\rm AW}\rangle_J}
{\langle\gamma\rangle_J\langle Q_{\rm AW}\rangle_J}
=
1+
\frac{{\rm Cov}_J(\gamma,Q_{\rm AW})}
{M_1\langle Q_{\rm AW}\rangle_J}.
\label{6.awfac}
\end{equation}
Exact factorization would give $\mathcal F_{\rm AW}=1$. 
Using the Maxwell-J\"uttner integral in Eq. \eqref{3.4}, the two limiting forms are
\begin{equation}
\mathcal F_{\rm AW}(\zeta)
=
1+
\frac{2}{\zeta}
+
\mathcal{O}(\zeta^{-2}),
\qquad
\zeta\gg1,
\label{6.awfacnr}
\end{equation}
and
\begin{equation}
\mathcal F_{\rm AW}(\zeta)
=
\frac{4}{3}
+
\mathcal{O}(\zeta^2),
\qquad
\zeta\ll1.
\label{6.awfacur}
\end{equation}
The nonrelativistic correction vanishes with temperature, while the ultrarelativistic factorization error remains finite.

We can also test the finite central-moment expansion directly. 
For the function in Eq. \eqref{6.awq},
\begin{equation}
Q'_{\rm AW}(\gamma)
=
1+2\gamma^{-2}-3\gamma^{-4},
\qquad
Q''_{\rm AW}(\gamma)
=
-4\gamma^{-3}+12\gamma^{-5}.
\label{6.awder}
\end{equation}
Keeping the first covariance term and then the next term gives
\begin{equation}
\mathcal F_{\rm AW}^{[1]}
=
1+
\frac{Q'_{\rm AW}(M_1)\mu_2}
{M_1\langle Q_{\rm AW}\rangle_J},
\qquad
\mathcal F_{\rm AW}^{[2]}
=
1+
\frac{Q'_{\rm AW}(M_1)\mu_2+\tfrac12Q''_{\rm AW}(M_1)\mu_3}
{M_1\langle Q_{\rm AW}\rangle_J}.
\label{6.awtrunc}
\end{equation}
The left panel of Fig. \ref{fig_r2_quantitative_examples} evaluates Eqs. \eqref{6.awfac} and \eqref{6.awtrunc} without adjustable coefficients. 
The exact ratio departs from unity as the gas becomes relativistic and tends to $4/3$. 
The second truncation is not uniformly closer to the exact curve than the first one. 
We therefore use the exact ratio in quantitative statements and regard Eq. \eqref{4.18} only as a local expansion when the higher central moments are small.

The nonaffine formula can be made to reproduce an RTA example only after the memory kernel is chosen appropriately. 
Equating the minimally averaged nonaffine result to Eq. \eqref{6.7} gives the effective memory amplitude required by an RTA-matched closure,
\begin{equation}
\tilde{\nu}_{0}^{\rm RTA}(T)
=
\frac{
10\pi^2m c_s^3
}{
\kappa R_0^2\omega_D
}
\,
\frac{
n\tau_R m\,\mathcal R_\eta(\zeta)
}{
F_0(\zeta)
}.
\label{6.11}
\end{equation}
Thus the kinetic relaxation time and the nonaffine zero-frequency memory kernel are not identical objects. 
They become related only after specifying the collective spectrum and the nonaffine force correlator. 
This observation is the main lesson of the RTA example.

\subsection{Proof-of-principle finite-memory bath}
\label{subsec_finite_memory_bath}
We next specify a solvable oscillator bath so that the frequency dependence is no longer left arbitrary. 
The generalized Langevin oscillator-bath formalism represents the friction memory function as a cosine transform of a positive bath spectral weight \cite{Zwanzig_1973,Zaccone_2023,Zaccone_2024b}. 
We write
\begin{equation}
\nu(t)
=
\int_0^\infty
\mathcal J_\nu(\Omega)\cos(\Omega t)\,d\Omega .
\label{6.memcos}
\end{equation}
For a Lorentzian bath spectrum,
\begin{equation}
\mathcal J_\nu(\Omega)
=
\frac{2\nu_0}{\pi\left(1+\Omega^2\tau_m^2\right)},
\label{6.lorentzbath}
\end{equation}
Eq. \eqref{6.memcos} gives
\begin{equation}
\nu_D(t)
=
\frac{\nu_0}{\tau_m}e^{-t/\tau_m}\Theta(t),
\qquad
\widetilde\nu_D(\omega)
=
\frac{\nu_0}{1+i\omega\tau_m},
\label{6.expmem}
\end{equation}
where the transform uses the Fourier convention of Sec. \ref{sec_rel_nonaffine}. 
The short-memory limit $\tau_m\to0$ gives $\widetilde\nu_D(\omega)\to\nu_0$ at fixed $\omega$ and recovers frequency-independent friction.

For a compact proof of principle, we retain one stable collective mode of frequency $\omega_0$ and affine weight $\mathcal A_0=\Xi_{0,xy}^2/\mathring V$. 
Within the scalar branch we set $\bar\gamma=M_1(\zeta)$. 
Equation \eqref{2.8} then gives
\begin{equation}
G_D^*(\omega)
=
\frac{\mathcal A_0}
{m\bar\gamma\omega^2-m\omega_0^2-i\bar\gamma\omega\nu_0/(1+i\omega\tau_m)}.
\label{6.gdrude}
\end{equation}
The corresponding decay exponents $z$ follow from the Laplace-domain denominator and satisfy
\begin{equation}
m\bar\gamma\tau_m z^3
+
m\bar\gamma z^2
+
\left(m\omega_0^2\tau_m+\bar\gamma\nu_0\right)z
+
m\omega_0^2
=
0.
\label{6.memcubic}
\end{equation}
At $\tau_m=0$, Eq. \eqref{6.memcubic} reduces to the familiar quadratic damped-mode equation. 
A finite $\tau_m$ adds a third decay scale associated with the bath memory. 
As $\tau_m$ decreases, that extra pole moves to short times while the other two poles approach the frequency-independent-friction result.

\begin{figure}
\centering
\includegraphics[width=0.48\textwidth]{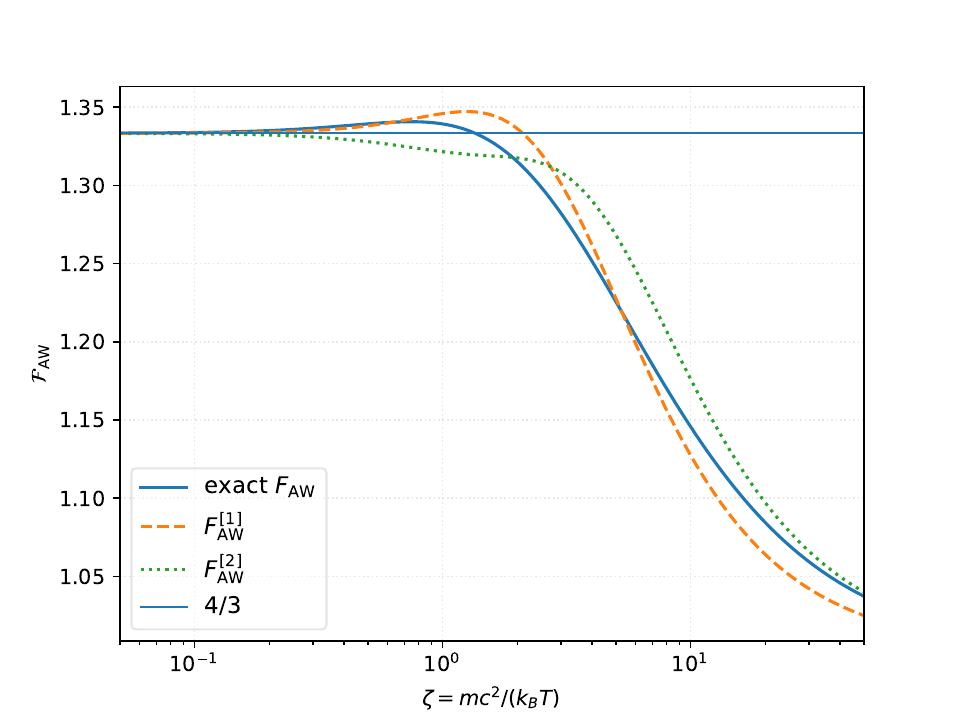}
\includegraphics[width=0.48\textwidth]{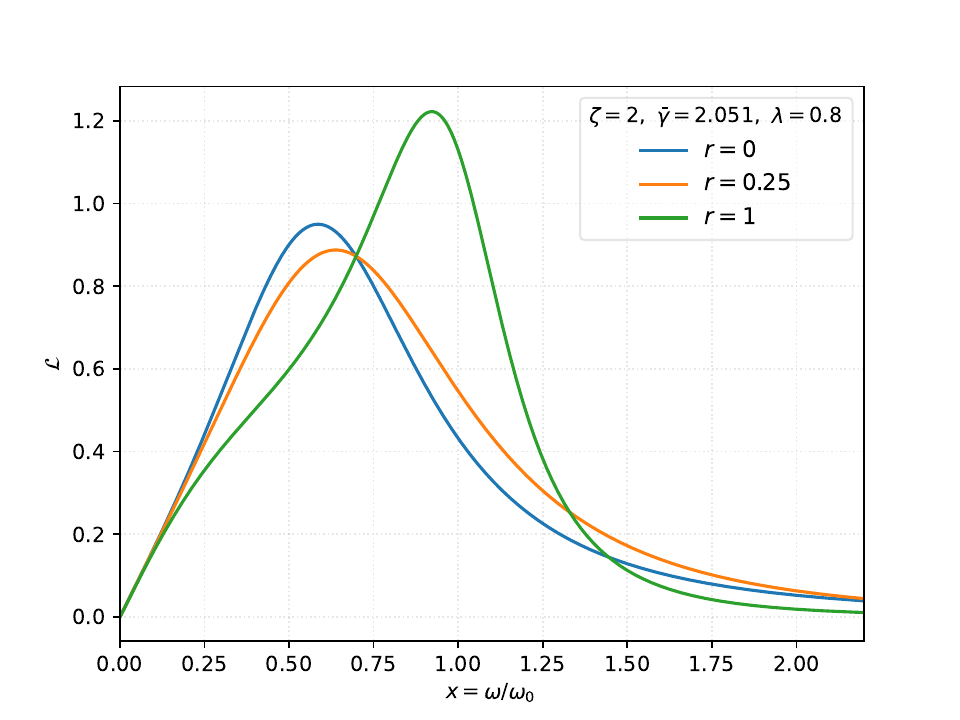}
\caption{Quantitative tests of momentum factorization and finite temporal memory. 
The left panel compares the exact Anderson-Witting factorization ratio in Eq. \eqref{6.awfac} with the two finite central-moment truncations in Eq. \eqref{6.awtrunc}. 
No adjustable covariance coefficients enter this calculation. 
The right panel shows the normalized single-mode loss response in Eq. \eqref{6.memloss}. 
We use $\zeta=2$, which gives $\bar\gamma=M_1(2)\simeq2.051$, together with $\lambda=0.8$ and $r=0,0.25,1$. 
The curve at $r=0$ is the short-memory result, while finite $r$ shifts and reshapes the dissipative peak.}
\label{fig_r2_quantitative_examples}
\end{figure}

The change in the loss spectrum is explicit after defining
\begin{equation}
x=\frac{\omega}{\omega_0},
\qquad
r=\omega_0\tau_m,
\qquad
\lambda=\frac{\nu_0}{m\omega_0}.
\label{6.memdimless}
\end{equation}
Writing $\mathcal L=m\omega_0^2G''/\mathcal A_0$, we obtain
\begin{equation}
\mathcal L(x)
=
\frac{b(x)}{a^2(x)+b^2(x)},
\qquad
b(x)
=
\frac{\bar\gamma\lambda x}{1+r^2x^2},
\qquad
a(x)
=
1-\bar\gamma x^2
+
\frac{\bar\gamma\lambda r x^2}{1+r^2x^2}.
\label{6.memloss}
\end{equation}
The right panel of Fig. \ref{fig_r2_quantitative_examples} shows Eq. \eqref{6.memloss} for $\zeta=2$, $\lambda=0.8$, and three values of $r$. 
Finite memory shifts the dissipative peak and changes its width and height. 
The zero-frequency viscosity is unchanged in this example because $\widetilde\nu_D(0)=\nu_0$ for every $\tau_m$. 
Hence the finite-memory effect appears in the frequency-dependent response and in the additional decay pole rather than in the static shear coefficient.

\subsection{Massive-gas and hard-sphere examples}
\label{subsec_massive_hardsphere_examples}
Massive relativistic gases provide a useful test of the averaged nonaffine construction because all dependence on $m/T$ is explicit. 
In moment-method and relaxation-time calculations, the shear viscosity is controlled by tensorial momentum moments of the equilibrium distribution \cite{Denicol:2012cn,Ambrus:2023qcl}. 

Existing kinetic treatments of finite-mass and quasiparticle systems place the present fixed-mass comparison in a broader setting.
Plumari et al. compared Green-Kubo, Chapman-Enskog, and relaxation-time estimates for both massive and massless relativistic particles and for isotropic and anisotropic scattering.
They found that a simple relaxation-time approximation can substantially underestimate the shear viscosity when the angular dependence of scattering is important \cite{Plumari:2012ep}.
Alqahtani, Nopoush, and Strickland introduced a temperature-dependent quasiparticle mass fitted to lattice-QCD thermodynamics in anisotropic hydrodynamics.
Their equations follow from moments of a Boltzmann equation with a relaxation-time collision term and include the additional thermodynamic contribution required by the varying quasiparticle mass \cite{Alqahtani:2015qja}.
Parisi et al. treated relativistic binary mixtures in Chapman-Enskog theory and compared the two-species result with a Green-Kubo evaluation of Boltzmann evolution for a quasiparticle description of QCD thermodynamics \cite{Parisi:2025gwq}.
Their results show that interspecies scattering can change the temperature dependence of the shear viscosity relative to combinations of isolated one-species viscosities.
Our present model assumes one classical species with fixed rest mass.
The function $F_0(\zeta)$ therefore isolates the equilibrium Lorentz-factor contribution associated with that fixed mass.
A temperature-dependent effective mass would make $\zeta=m(T)/T$ state dependent and would require a thermodynamically consistent extension of the equilibrium sector.
Species-dependent and angular scattering effects would instead enter the momentum-dependent dissipative response.
We do not include these extensions here, so the comparison below should be read as a fixed-mass one-species reference rather than as a quasiparticle or anisotropic-hydrodynamic description.

The normalized kinetic shear weight in Eq. \eqref{6.8} is one such moment. 
The minimally averaged nonaffine theory instead contains
\begin{equation}
F_0(\zeta)
=
\left\langle\gamma\right\rangle_J .
\label{6.12}
\end{equation}
Consequently, the ratio of the kinetic shear weight to the minimal nonaffine Lorentz enhancement is
\begin{equation}
\mathcal C_\eta(\zeta)
=
\frac{\mathcal R_\eta(\zeta)}{F_0(\zeta)}
=
\frac{
\zeta
\left\langle
(\gamma^2-1)^2/\gamma^2
\right\rangle_J
}{
15\langle\gamma\rangle_J
}.
\label{6.13}
\end{equation}
This ratio measures the amount of momentum selectivity that the nonaffine memory kernel must carry if the averaged nonaffine formula is to imitate a constant-$\tau_R$ kinetic model. 
Using the limiting results above,
\begin{equation}
\mathcal C_\eta(\zeta)
=
\frac{1}{\zeta}
+
\mathcal{O}(\zeta^{-2}),
\qquad
\zeta\gg1,
\label{6.14}
\end{equation}
and
\begin{equation}
\mathcal C_\eta(\zeta)
=
\frac{4}{15}
+
\mathcal{O}(\zeta^2\ln\zeta),
\qquad
\zeta\ll1.
\label{6.15}
\end{equation}
The nonrelativistic behavior in Eq. \eqref{6.14} shows that a kinetic relaxation-time closure and a pure Lorentz-enhancement closure distribute temperature dependence differently. 
In the kinetic expression, the shear tensor itself contributes powers of momentum. 
In the nonaffine expression, the same temperature dependence must be carried by the collective factors $\kappa R_0^2$, $\omega_D$, $c_s$, and $\tilde{\nu}_0$.

\begin{figure}
\centering
\includegraphics[width=0.48\textwidth]{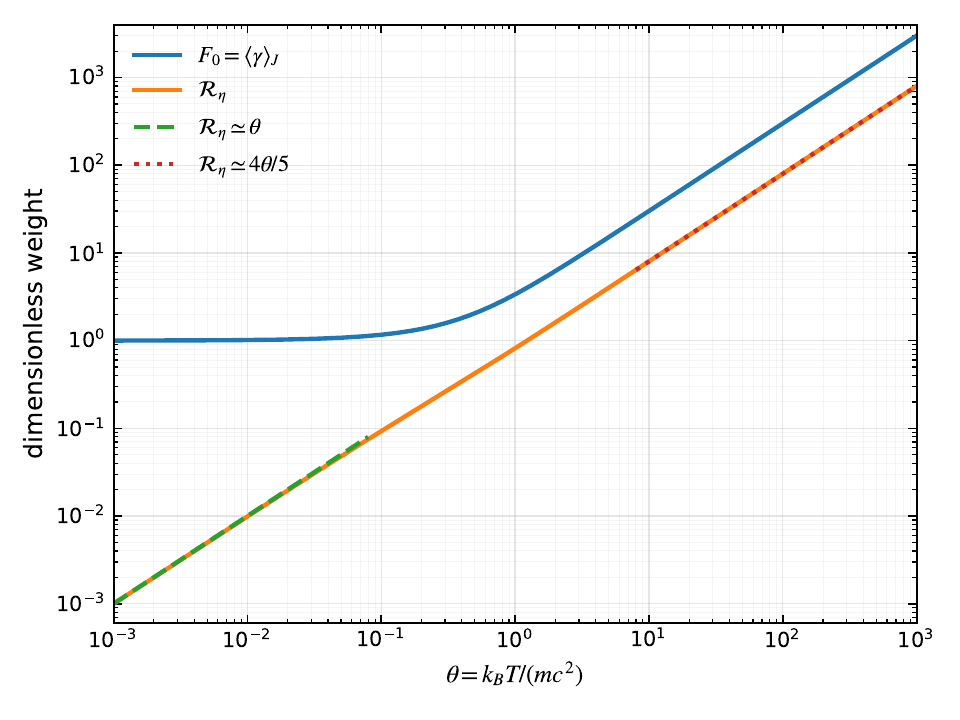}
\includegraphics[width=0.48\textwidth]{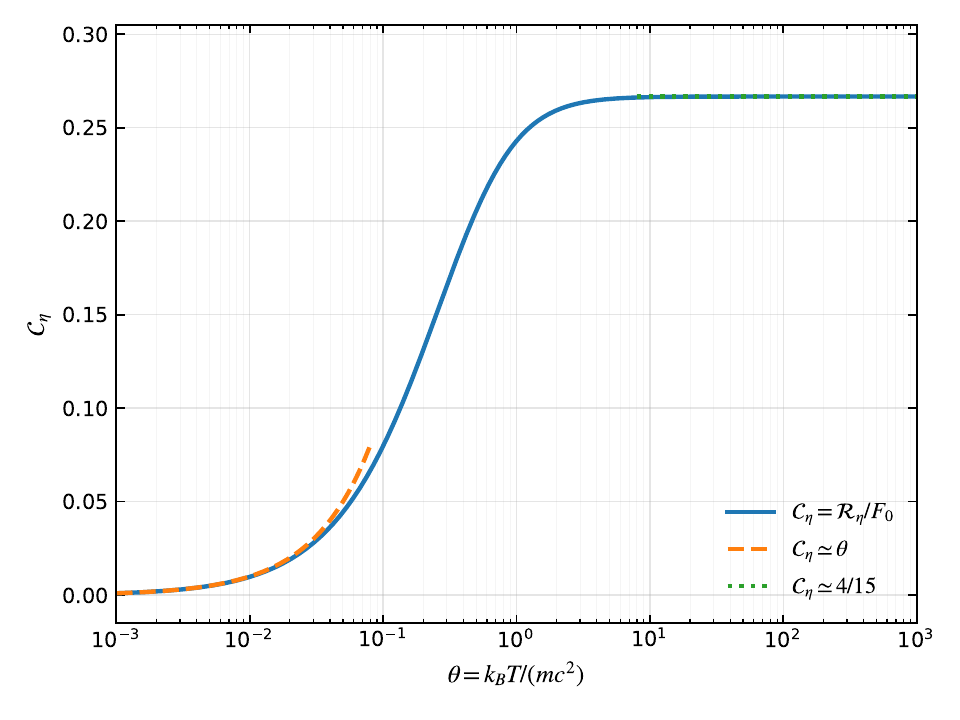}
\caption{
Comparison between the kinetic relaxation-time shear weight and the minimal
nonaffine Lorentz enhancement.  The left panel shows
$F_0(\zeta)=\langle\gamma\rangle_J$ and the normalized RTA shear weight
$\mathcal R_\eta(\zeta)$, together with the nonrelativistic and ultrarelativistic
limits of $\mathcal R_\eta$.  The right panel shows
$\mathcal C_\eta(\zeta)=\mathcal R_\eta(\zeta)/F_0(\zeta)$, the momentum
selectivity that an effective nonaffine memory amplitude must carry in order to
imitate a constant-$\tau_R$ kinetic closure.
}
\label{fig_rta-nonaffine-comparison}
\end{figure}
Figure \ref{fig_rta-nonaffine-comparison} makes the distinction between the
minimal nonaffine average and the kinetic shear tensor moment explicit.  The
minimal nonaffine closure contains the scalar Lorentz moment
$F_0=\langle\gamma\rangle_J$, whereas the RTA shear viscosity is controlled by
the tensorial momentum weight $(\gamma^2-1)^2/\gamma^2$.  Consequently,
$\mathcal C_\eta$ tends to $\theta$ in the nonrelativistic regime and to
$4/15$ in the ultrarelativistic regime.  This behavior shows that a kinetic
relaxation time and a nonaffine zero-frequency memory kernel are not identical
objects.  They can be related only after the collective spectral factors and the
momentum dependence of the effective memory amplitude have been specified.

The hard-sphere ultrarelativistic gas illustrates the same point in a sharper way. 
For a dilute gas with an approximately temperature-independent microscopic cross section, kinetic-theory examples generally produce a viscosity proportional to a pressure times a relaxation time, $\eta\sim P\tau$, with the relaxation time fixed by the mean free path and collision rate \cite{Anderson:1974nyl,Denicol:2019lio}. 
If the number density scales as $n\sim T^3$ and the cross section is constant, then $\tau\sim 1/(n\sigma)$ and the kinetic estimate scales as
\begin{equation}
\eta_{\rm dilute}
\sim
P\tau
\sim
T^4\frac{1}{T^3\sigma}
\sim
\frac{T}{\sigma}.
\label{6.16}
\end{equation}
By contrast, the dense-fluid nonaffine scaling derived in Sec. \ref{subsec_ultrarelativistic_limit} gives
\begin{equation}
\eta_J^{(0)}
\sim
\tilde{\nu}_\infty(T)T^3,
\label{6.17}
\end{equation}
after the thermal stiffness and collective cutoff have been specified as in Eq. \eqref{5.18}. 
Equations \eqref{6.16} and \eqref{6.17} need not agree because they describe different physical closures. 
The first is a dilute binary-collision estimate with a specified cross section. 
The second is a dense collective-response estimate in which the zero-frequency memory kernel carries the unresolved collisional and many-body information. 
The comparison therefore identifies the precise burden placed on $\tilde{\nu}_\infty(T)$: it must encode whether the gas behaves as a dilute kinetic system, a dense correlated fluid, or an intermediate system.

This distinction is also relevant for hydrodynamic attractor examples. 
In the ultrarelativistic hard-sphere system studied by Denicol and Noronha, the microscopic dynamics is governed by the Boltzmann equation with a nonlinear collision kernel, and the hydrodynamic behavior depends on the separation between microscopic and macroscopic scales \cite{Denicol:2019lio}. 
The averaged nonaffine formula does not reproduce such an attractor by itself. 
It can only supply the shear viscosity entering a hydrodynamic closure after its memory kernel has been fixed by the same microscopic physics that determines the kinetic collision operator.

\subsection{Relation to Kubo response}
\label{subsec_kubo_interpretation}

We compare the causal nonaffine shear response with the Green-Kubo definition. 
For a relativistic fluid in equilibrium, the shear viscosity may be defined through the retarded stress correlator in the shear channel,
\begin{equation}
\eta
=
\lim_{\omega\to0}
\frac{1}{\omega}
{\rm Im}\,
G^R_{T^{xy}T^{xy}}(\omega,\mathbf 0),
\label{6.18}
\end{equation}
where $G^R_{T^{xy}T^{xy}}$ is the retarded two-point function of the microscopic stress tensor \cite{Kubo_1957,Denicol:2012cn}. 
The object in Eq. \eqref{6.18} is therefore an operator correlation function of the underlying microscopic theory.

The nonaffine calculation starts from a different set of microscopic variables. 
Extending the fixed-momentum modal denominator in Eq. \eqref{2.8} to the thermal average that underlies Eq. \eqref{4.1}, we define its causal shear modulus by
\begin{equation}
G^R_{\rm NA}(\omega)
=
\frac{1}{\mathring V}
\int_{\mathbb R^3}d^3p\,f_J(\mathbf p,\zeta)
\int_0^{\omega_D}d\omega_p\,
\frac{
g(\omega_p)\Gamma_{xyxy}(\omega_p,\mathbf p)
}{
m\gamma(\mathbf p)\omega^2
-
m\omega_p^2
-
i\gamma(\mathbf p)\tilde{\nu}(\omega,\mathbf p,\omega_p)\omega
}.
\label{6.19}
\end{equation}
Its imaginary part reproduces the thermally averaged loss response used in Sec. \ref{sec_averaged_viscosity}. 
Equation \eqref{6.19} therefore gives the causal response generated by the nonaffine model. 
We do not identify $G^R_{\rm NA}$ with $G^R_{T^{xy}T^{xy}}$ on this basis alone. 
Establishing equality would require deriving the microscopic stress tensor within the same particle-bath model and evaluating its retarded equilibrium correlation function.

For a fluid, the static affine and nonaffine elastic parts cancel in the zero-frequency shear modulus \cite{Zaccone_2023}. 
We isolate the dissipative part of Eq. \eqref{6.19} by defining the complex shear viscosity
\begin{equation}
\eta^*_{\rm NA}(\omega)
=
\frac{
G^R_{\rm NA}(\omega)-G^R_{\rm NA}(0)
}{i\omega}.
\label{6.complexeta}
\end{equation}
This local definition is sufficient for the shear channel because the subtraction removes the static elastic contribution while leaving the linear dissipative term unchanged. 
The zero-frequency coefficient is
\begin{equation}
\eta_J
=
\lim_{\omega\to0}
{\rm Re}\,\eta^*_{\rm NA}(\omega),
\label{6.etazero}
\end{equation}
which reproduces Eq. \eqref{4.3}. 
Causality then defines a time-domain shear-response function $\mathcal R_\eta(t)$ through
\begin{equation}
\eta^*_{\rm NA}(\omega)
=
\int_0^\infty
\mathcal R_\eta(t)e^{i\omega t}\,dt,
\qquad
\mathcal R_\eta(t<0)=0.
\label{6.shearresponse}
\end{equation}
The function $\mathcal R_\eta$ is obtained only after the momentum and collective-mode integrations in Eq. \eqref{6.19}. 
It is therefore not the same object as the microscopic friction function $\tilde{\nu}$. 
The latter enters each mode before the thermal and spectral integrations, while $\mathcal R_\eta$ describes the resulting macroscopic shear response.

The zero-frequency positivity statement remains direct. 
If $\tilde{\nu}(0,\mathbf p,\omega_p)\geq0$ and $\Gamma_{xyxy}(\omega_p,\mathbf p)\geq0$, Eq. \eqref{4.3} gives
\begin{equation}
\eta_J\geq0.
\label{6.20}
\end{equation}
Finite-frequency passivity requires the stronger condition ${\rm Re}\,\eta^*_{\rm NA}(\omega)\geq0$ for real $\omega$. 
The structural similarity between Eqs. \eqref{6.18} and \eqref{6.19} thus supplies a comparison framework, but it does not prove equality of the two retarded functions. 
A quantitative Kubo test must compute both responses for the same microscopic system.

\section{Hydrodynamic Consistency and Causality}
\label{sec_hydro_consistency}
\subsection{Frame-independent transport coefficient versus frame-dependent constitutive relation}
\label{subsec_transport_vs_frame}

The Maxwell-J\"uttner averaged theory supplies a local-rest-frame shear coefficient and a frequency-dependent shear response. 
A relativistic evolution problem still requires a constitutive energy-momentum tensor. 
For a neutral fluid, we write

\begin{equation}
T^{\mu\nu}
=
\varepsilon u^\mu u^\nu
-
P\Delta^{\mu\nu}
+
\pi^{\mu\nu}
+
\mathcal Q^\mu u^\nu
+
\mathcal Q^\nu u^\mu
+
\mathcal A u^\mu u^\nu
-
\mathcal B\Delta^{\mu\nu},
\label{7.1}
\end{equation}

Here $\Delta^{\mu\nu}=g^{\mu\nu}-u^\mu u^\nu$, while $\pi^{\mu\nu}$ is transverse and traceless. 
The quantities $\mathcal A$, $\mathcal B$, and $\mathcal Q^\mu$ represent frame-dependent first-order terms \cite{Kovtun:2019hdm,Bemfica:2019knx,Rocha:2023hts}. 
The transverse traceless channel isolates $\eta_J$ at linear order around homogeneous equilibrium.

The shear tensor is
\begin{equation}
\sigma^{\mu\nu}
=
\Delta^{\mu\nu\alpha\beta}\partial_\alpha u_\beta,
\qquad
\Delta^{\mu\nu\alpha\beta}
=
\frac{1}{2}
\left(
\Delta^{\mu\alpha}\Delta^{\nu\beta}
+
\Delta^{\mu\beta}\Delta^{\nu\alpha}
\right)
-
\frac{1}{3}\Delta^{\mu\nu}\Delta^{\alpha\beta}.
\label{7.2}
\end{equation}
The first-order shear constitutive relation is then
\begin{equation}
\pi^{\mu\nu}
=
2\eta_J\,\sigma^{\mu\nu}
+
\mathcal{O}(\partial^2),
\label{7.3}
\end{equation}
with the sign convention fixed by the definition of $\sigma^{\mu\nu}$ and by the mostly-minus metric. 
The coefficient $\eta_J$ is the scalar computed in Sec. \ref{sec_averaged_viscosity}. 
In the local rest frame, choose a transverse contravariant velocity perturbation $v_y(t,x)\equiv\delta u^y(t,x)$ with wave vector along $x$. 
Since $u_y=-v_y$ to linear order in the mostly-minus metric, one has $\partial_\mu\delta u^\mu=0$, scalar corrections are absent, and Eq. \eqref{7.2} gives
\begin{equation}
\sigma^{xy}
=
-\frac{1}{2}\partial_x v_y .
\label{7.4}
\end{equation}
Consequently, Eq. \eqref{7.3} gives
\begin{equation}
\delta T^{xy}
=
\pi^{xy}
=
-\eta_J\,\partial_x v_y .
\label{7.5}
\end{equation}

Equation \eqref{7.5} identifies $\eta_J$ with the transverse traceless stress response. 
The remaining scalar and vector sectors and the frame choice are separate and determine whether the complete evolution equations are causal and stable \cite{Kovtun:2019hdm,Bemfica:2019knx,Ambrus:2023qcl}. 
The present theory therefore supplies the shear coefficient and its microscopic frequency dependence rather than a complete relativistic fluid theory.

\subsection{Microscopic shear response and transient relaxation}
\label{subsec_memory_transient_relaxation}

The frequency-dependent nonaffine response in Eq. \eqref{6.complexeta} already determines the causal shear-response function in Eq. \eqref{6.shearresponse}. 
We therefore do not introduce an independent hydrodynamic response function. 
In the shear sector, the corresponding time-domain relation is
\begin{equation}
\pi^{\mu\nu}(t)
=
2\int_{-\infty}^{t}
\mathcal R_\eta(t-t')\,
\sigma^{\mu\nu}(t')\,dt'.
\label{7.6}
\end{equation}
Its zero-frequency value gives
\begin{equation}
\eta_J
=
\int_0^\infty \mathcal R_\eta(t)\,dt .
\label{7.7}
\end{equation}
Fourier transformation of Eq. \eqref{7.6} gives
\begin{equation}
\tilde{\pi}^{\mu\nu}(\omega)
=
2\eta^*_{\rm NA}(\omega)\tilde{\sigma}^{\mu\nu}(\omega).
\label{7.8}
\end{equation}
If the first two time moments exist, expansion of Eq. \eqref{6.shearresponse} around $\omega=0$ yields
\begin{equation}
\eta^*_{\rm NA}(\omega)
=
\eta_J
+
i\omega\eta_J\tau_\eta^{(1)}
-
\frac{\omega^2}{2}
\int_0^\infty t^2\mathcal R_\eta(t)\,dt
+
\mathcal{O}(\omega^3),
\label{7.9}
\end{equation}
where we define the first-moment time scale
\begin{equation}
\tau_\eta^{(1)}
=
\frac{1}{\eta_J}
\int_0^\infty t\,\mathcal R_\eta(t)\,dt .
\label{7.10}
\end{equation}
Equation \eqref{7.10} is the coefficient of the first low-frequency correction. 
For a general causal response, it need not equal a physical decay time. 
The physical relaxation scales follow from the poles and other nonanalytic structures of $\eta^*_{\rm NA}(\omega)$ in the complex-frequency plane. 
The Lorentzian oscillator-bath example makes this distinction explicit because Eq. \eqref{6.memcubic} contains three decay poles when $\tau_m$ is finite.

To compare with a single-relaxation-time hydrodynamic theory, we now specialize to a one-pole response. 
We take
\begin{equation}
\mathcal R_\eta(t)
=
\frac{\eta_J}{\tau_\pi}e^{-t/\tau_\pi}\Theta(t),
\qquad
\tau_\pi=\tau_\eta^{(1)}.
\label{7.11}
\end{equation}
Substituting Eq. \eqref{7.11} into Eq. \eqref{7.6} and differentiating gives
\begin{equation}
\tau_\pi\frac{d\pi^{\mu\nu}}{dt}
+
\pi^{\mu\nu}
=
2\eta_J\sigma^{\mu\nu}.
\label{7.12}
\end{equation}
In a covariant hydrodynamic equation, we replace the ordinary derivative by the comoving derivative and project the result so that $\pi^{\mu\nu}$ remains transverse and traceless. 
The leading shear equation is
\begin{equation}
\tau_\pi\Delta^{\mu\nu}_{\alpha\beta}D\pi^{\alpha\beta}
+
\pi^{\mu\nu}
=
2\eta_J\sigma^{\mu\nu}
+
\mathcal{O}(\partial^2),
\label{7.13}
\end{equation}
which has the same structural role as the shear sector of transient second-order relativistic hydrodynamics \cite{Denicol:2012cn,Ambrus:2023qcl}. 
For the exponential response, Eq. \eqref{6.shearresponse} gives
\begin{equation}
\eta^*_{\rm NA}(\omega)
=
\frac{\eta_J}{1-i\omega\tau_\pi}.
\label{7.14}
\end{equation}
The pole at $\omega=-i/\tau_\pi$ shows why $\tau_\pi$ is a physical relaxation time in this special case. 
The instantaneous Navier-Stokes relation follows as $\tau_\pi\to0$ at fixed $\eta_J$. 
For a response with several poles, one generally needs several relaxation variables or the full temporal response rather than Eq. \eqref{7.12}.

\subsection{Stability constraints}
\label{subsec_stability_constraints}
 The stability and causality analysis below refers to the one-pole specialization in Eq. \eqref{7.11}. A response with several poles requires the corresponding enlarged transient system or the full temporal relation in Eq. \eqref{7.6}.
The first consistency requirement is positivity of entropy production in the shear channel. 
For the instantaneous limit, the local quadratic entropy production rate is proportional to $\pi^{\mu\nu}\sigma_{\mu\nu}/T$. 
Using Eq. \eqref{7.3}, one obtains
\begin{equation}
\nabla_\mu S^\mu_{\rm sh}
=
\frac{2\eta_J}{T}
\sigma^{\mu\nu}\sigma_{\mu\nu}
+
\mathcal{O}(\partial^3).
\label{7.15}
\end{equation}
Thus the microscopic condition $\eta_J\geq0$ is necessary for nonnegative shear entropy production. 
From Eq. \eqref{4.3}, this follows if
\begin{equation}
\tilde{\nu}(0,\mathbf p,\omega_p)\geq0,
\qquad
\Gamma_{xyxy}(\omega_p,\mathbf p)\geq0,
\label{7.16}
\end{equation}
because the Maxwell-J\"uttner weight, the mode density, and $\gamma(\mathbf p)$ are nonnegative. 
The first inequality is passivity of the bath, while the second is positivity of the affine-force spectral weight in the shear channel.

For the causal shear response, finite-frequency passivity requires a positive quadratic dissipation functional. 
For any smooth test shear history $X_{\mu\nu}(t)$ with compact support, we require
\begin{equation}
\int_{-\infty}^{\infty}dt
\int_{-\infty}^{\infty}dt'\,
X_{\mu\nu}(t)\,
\mathcal R_\eta(t-t')\,
X^{\mu\nu}(t')
\geq0 .
\label{7.17}
\end{equation}
In frequency space this requires
\begin{equation}
{\rm Re}\,\eta^*_{\rm NA}(\omega)\geq0
\qquad
\hbox{for all real } \omega .
\label{7.18}
\end{equation}
For the exponential response in Eq. \eqref{7.11}, Eq. \eqref{7.14} gives
\begin{equation}
{\rm Re}\,\eta^*_{\rm NA}(\omega)
=
\frac{\eta_J}{1+\omega^2\tau_\pi^2},
\label{7.19}
\end{equation}
so passivity requires $\eta_J\geq0$ and $\tau_\pi>0$.

A second consistency requirement is linear stability of transverse perturbations. 
Consider a homogeneous equilibrium state with enthalpy density $w=\varepsilon+P$ and a transverse contravariant velocity perturbation $v_y(t,x)=\delta u^y(t,x)$. 
Linear momentum conservation gives
\begin{equation}
w\,\partial_t v_y
+
\partial_x\pi^{xy}
=
0 .
\label{7.20}
\end{equation}
Using the relaxation equation \eqref{7.12} in the local rest frame and $\sigma^{xy}=-\partial_x v_y/2$, one has
\begin{equation}
\tau_\pi\partial_t\pi^{xy}
+
\pi^{xy}
=
-\eta_J\partial_x v_y .
\label{7.21}
\end{equation}
Eliminating $\pi^{xy}$ between Eqs. \eqref{7.20} and \eqref{7.21} yields
\begin{equation}
\tau_\pi\partial_t^2 v_y
+
\partial_t v_y
-
\frac{\eta_J}{w}\partial_x^2 v_y
=
0 .
\label{7.22}
\end{equation}
For a Fourier perturbation $v_y\propto e^{-i\omega t+ikx}$, the dispersion relation is
\begin{equation}
\tau_\pi\omega^2
+
i\omega
-
\frac{\eta_J}{w}k^2
=
0 .
\label{7.23}
\end{equation}
At small $k$, the hydrodynamic root is
\begin{equation}
\omega
=
-i\frac{\eta_J}{w}k^2
+
\mathcal{O}(k^4),
\label{7.24}
\end{equation}
which is stable when $\eta_J/w>0$. 
The nonhydrodynamic root is
\begin{equation}
\omega
=
-\frac{i}{\tau_\pi}
+
i\frac{\eta_J}{w}k^2
+
\mathcal{O}(k^4),
\label{7.25}
\end{equation}
which is damped at sufficiently small $k$ when $\tau_\pi>0$. 
At large $k$, Eq. \eqref{7.22} has characteristic shear speed
\begin{equation}
v_T^2
=
\frac{\eta_J}{w\tau_\pi}.
\label{7.26}
\end{equation}
A necessary transverse causality condition is therefore
\begin{equation}
0
\leq
\frac{\eta_J}{w\tau_\pi}
\leq
1 .
\label{7.27}
\end{equation}
This condition is not a full proof of nonlinear causality for the complete relativistic fluid. 
 It is the shear-sector condition implied by the one-pole transverse mode. 
The complete causal and stable theory additionally depends on the scalar sector, frame choice, equation of state, and possible charge diffusion terms \cite{Kovtun:2019hdm,Bemfica:2019knx,Rocha:2023hts,Ambrus:2023qcl}. 
 Nevertheless, Eq.~\eqref{7.27} gives a direct microscopic target for the one-pole reduction. Once $\eta_J$ and the dominant pole time $\tau_\pi$ are extracted from the same microscopic shear response, their ratio must be checked against the enthalpy density before the result is used in relativistic hydrodynamic simulations.

\begin{figure}
\centering
\includegraphics[width=0.80\linewidth]{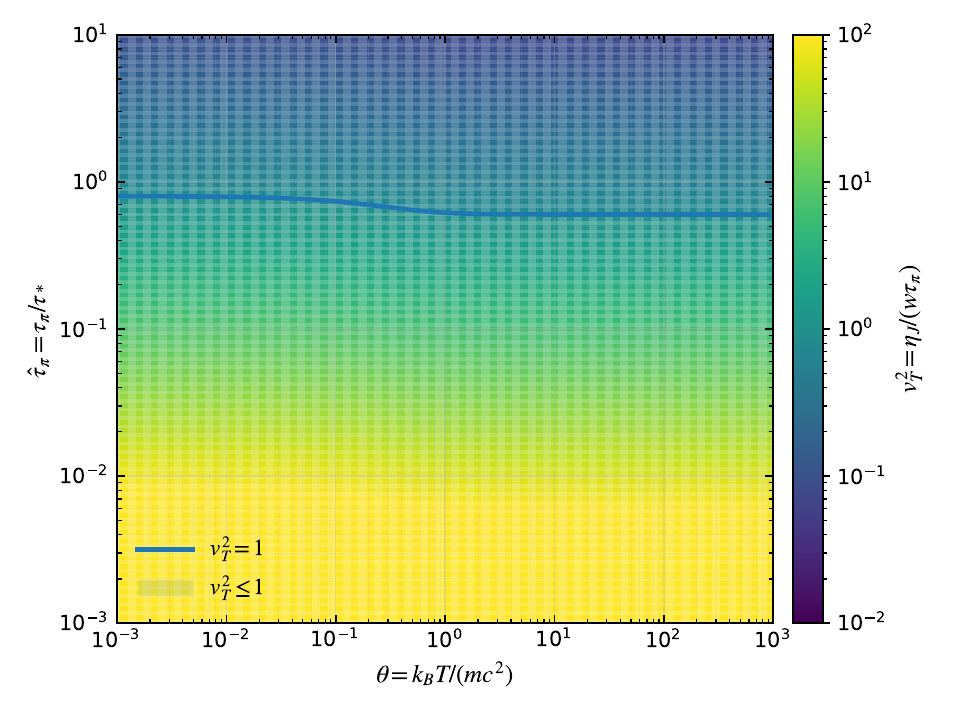}
\caption{
Shear-sector causality target for the one-pole transverse mode.
The color scale shows $v_T^2=\eta_J/(w\tau_\pi)$ in a minimal illustrative
closure with $\eta_J=\eta_{\rm NR}F_0(\theta)$ and
$w/(nmc^2)=F_0(\theta)+\theta$.  The solid curve is the boundary $v_T^2=1$.
The shaded region above it satisfies the necessary transverse causality condition
$v_T^2\leq1$.  The normalization
$\lambda_\eta=\eta_{\rm NR}/(nmc^2\tau_*)$ is fixed only for visualization, so
the figure applies only after the microscopic shear response has been reduced to one dominant relaxation pole.
The plot is illustrative and is not a quantitative prediction for a specified microscopic fluid.
}
\label{fig_shear-causality-target}
\end{figure}
Figure \ref{fig_shear-causality-target} translates the transverse-mode condition
$v_T^2=\eta_J/(w\tau_\pi)\leq1$ into a relaxation-time target.  For the
illustrative ideal-gas enthalpy scaling used in the plot, the critical relaxation
time is
$\hat{\tau}_{\pi,\rm crit}=\lambda_\eta F_0/(F_0+\theta)$, where
$\hat{\tau}_\pi=\tau_\pi/\tau_*$ and
$\lambda_\eta=\eta_{\rm NR}/(nmc^2\tau_*)$.  The nonrelativistic limit gives
$\hat{\tau}_{\pi,\rm crit}\to\lambda_\eta$, while the ultrarelativistic limit
gives $\hat{\tau}_{\pi,\rm crit}\to3\lambda_\eta/4$.   The map emphasizes that, within the one-pole reduction,
the Maxwell-J\"uttner enhancement of the microscopic viscosity must be
accompanied by a sufficiently large pole relaxation time before the
coefficient can be inserted consistently into a transient relativistic shear
sector.

\section{Discussion}
\label{sec_discussion}
\subsection{What is genuinely new}
\label{subsec_genuinely_new}

The main advance is a finite-mass Maxwell-J\"uttner average of the relativistic nonaffine shear response. 
Within the scalar transverse mean-field branch, the minimal result is $F_0(\zeta)=\langle\gamma\rangle_J$. 
The general expression averages the full product of the Lorentz factor, friction response, and affine-force correlator. 
The Anderson-Witting shear weight provides a concrete test of this factorization. 
Its exact ratio approaches $4/3$ in the ultrarelativistic limit, while finite central-moment truncations lose reliability across the relativistic crossover. 
We therefore use the covariance series only as a local expansion when an exact momentum average is unavailable.

The finite-memory oscillator-bath example tests a second aspect of the framework. 
A nonzero bath memory time changes the frequency-dependent loss response and introduces an additional decay pole while preserving the same zero-frequency viscosity when $\nu_0$ is fixed. 
The first response moment $\tau_\eta^{(1)}$ then characterizes the low-frequency expansion, whereas physical relaxation times follow from the pole structure. 
Only a one-pole response permits the identification $\tau_\eta^{(1)}=\tau_\pi$ used in the transient shear equation.

\subsection{What remains model-dependent}
\label{subsec_model_dependent}

Quantitative predictions still require microscopic specification of the friction response, the affine-force correlator, and the admissible collective spectrum. 
The real-frequency Debye form used here applies to the positive Hessian branch of a mechanically stable short-time reference configuration. 
Negative Hessian eigenvalues and strongly overdamped excitations lie outside that reduction. 
For ideal hard spheres, the discontinuous interaction has no conventional Hessian at collision, so the hard-sphere calculation is only a scaling consistency test. 
These restrictions make the direct collective-mode interpretation most natural for dense or correlated fluids with transient local structure \cite{Lemaitre_2006,Zaccone_2011,Zaccone_2023}.

We also use the standard classical Maxwell-J\"uttner equilibrium measure. 
Alternative relativistic equilibrium measures would change the Lorentz moments, while quantum statistics, pair production, screening, and temperature-dependent quasiparticle masses would alter the microscopic shear response \cite{Dunkel:2006nk,Dunkel:2006hc,Rocha:2023hts,Ambrus:2023qcl}. 
Hydrodynamic consistency remains separate. 
The present calculation supplies the shear sector, while a complete causal and stable evolution theory must also specify the remaining constitutive sectors and the hydrodynamic frame \cite{Kovtun:2019hdm,Bemfica:2019knx,Rocha:2023hts}.

\subsection{Possible applications}
\label{subsec_possible_applications}

The finite-$\zeta$ formulas are most directly useful where $k_BT$ is comparable with $mc^2$ and neither asymptotic temperature limit is adequate. 
For dense relativistic fluids, the collective formulation can also retain finite temporal memory that a single relaxation-time approximation removes. 
A quantitative application requires the microscopic friction response and affine-force spectrum for the same system, followed by comparison with kinetic theory or microscopic simulation.

The most useful next calculations are therefore system-specific. 
One can derive the full frequency- and momentum-dependent friction response from a specified relativistic bath spectrum, compute the affine-force response for a dense relativistic fluid, and compare the resulting complex viscosity with massive relativistic kinetic theory. 
The pole structure can then be used to test the shear-sector causality condition in Eq.~\eqref{7.27} when a one-pole reduction is justified.

\section{Conclusion}
\label{sec_conclusion}
We have constructed a Maxwell-J\"uttner averaged extension of relativistic nonaffine shear-viscosity theory.  The present construction averages the fixed-$\gamma$ response only after the scalar transverse modal reduction. We have shown explicitly that the exact relativistic inertia is tensorial and that particle-dependent Lorentz factors can generate off-diagonal couplings in the Hessian basis. The scalar theory is therefore a mean-field approximation whose validity requires weak momentum-mode correlations and a predominantly transverse microscopic shear response. Within this stated branch, the minimal closure replaces the fixed Lorentz factor by its exact Maxwell-J\"uttner mean. In the general closure, the averaged object is the full product of the Lorentz factor, the zero-frequency memory kernel, and the nonaffine affine-force correlator. This distinction separates the purely kinematic part of relativistic enhancement from the dynamical part encoded in the bath and collective response.

The resulting theory gives a controlled interpolation between the nonrelativistic and ultrarelativistic regimes. In the nonrelativistic limit, the averaged enhancement approaches unity with analytic corrections in $k_B T/(mc^2)$, so the classical nonaffine viscosity formula is recovered.  Under the effective hard-sphere assumptions for the local stiffness scale, acoustic cutoff, sound speed, and memory amplitude, the usual square-root temperature law of the dilute nonrelativistic gas is reproduced as a scaling consistency test. It is not derived from a hard-sphere Hessian. In the ultrarelativistic limit, the Maxwell-J\"uttner enhancement contributes one additional power of temperature. The often-quoted cubic high-temperature scaling follows only after specifying the remaining microscopic scales, especially the memory kernel and the collective cutoff. Thus the cubic law is a conditional consequence of a dense-fluid closure, not a universal consequence of relativistic kinematics alone.

The comparison with relativistic kinetic-theory examples clarifies the domain of the construction. Relaxation-time kinetic theory weights the shear channel by tensorial momentum factors, whereas the minimal nonaffine closure weights the response by the Lorentz factor. Agreement between the two descriptions therefore requires a nontrivial momentum dependence in the memory kernel or in the nonaffine force correlator.  For the explicit Anderson-Witting shear weight, the exact Maxwell-J\"uttner calculation shows that factorization fails by a finite amount in the relativistic regime and tends to a ratio $4/3$ in the ultrarelativistic limit. A Lorentzian oscillator bath further gives a solvable finite-memory example in which the short-memory limit is recovered continuously and finite temporal memory produces an additional decay pole and a modified loss spectrum.  The Kubo comparison shows a structural similarity between the nonaffine causal response and the retarded stress correlator, but we do not identify the two without deriving the microscopic stress correlation in the same particle-bath theory. The complex nonaffine viscosity $\eta^*_{\rm NA}(\omega)$ instead gives the relevant relaxation spectrum directly. Its first moment defines the low-frequency scale $\tau_\eta^{(1)}$, while a physical shear relaxation time $\tau_\pi$ is obtained only in a one-pole reduction. Under that restriction, the transverse causality condition applies to $\eta_J/(w\tau_\pi)$.

Several extensions now follow naturally.  The most important is to derive the full momentum- and mode-dependent friction response from a specified relativistic bath spectral density, rather than retaining only its zero-frequency value. A second direction is to compute the affine-force correlator and the frequency-dependent shear response from microscopic simulations of relativistic classical gases or dense effective fluids. A third direction is to extend the present framework to bulk viscosity, where conformal-symmetry breaking and equation-of-state dependence should enter more directly. Finally, the averaged nonaffine construction should be tested against massive relativistic kinetic-theory calculations, hard-sphere gases, and QGP-inspired effective models, with careful attention to quantum statistics, pair production, screening, and hydrodynamic frame choice \cite{Anderson:1974nyl,Denicol:2012cn,Kovtun:2019hdm,Bemfica:2019knx,Denicol:2019lio,Rocha:2021zcw,Rocha:2023hts,Ambrus:2023qcl,Zaccone_2024}.

\acknowledgments
R. P. and A. \"O. would like to acknowledge networking support of the COST Action CA21106 - COSMIC WISPers in the Dark Universe: Theory, astrophysics and experiments (CosmicWISPers), the COST Action CA22113 - Fundamental challenges in theoretical physics (THEORY-CHALLENGES), the COST Action CA21136 - Addressing observational tensions in cosmology with systematics and fundamental physics (CosmoVerse), the COST Action CA23130 - Bridging high and low energies in search of quantum gravity (BridgeQG), and the COST Action CA23115 - Relativistic Quantum Information (RQI) funded by COST (European Cooperation in Science and Technology). A. \"O. also thanks to EMU, TUBITAK, ULAKBIM (Turkiye) and SCOAP3 (Switzerland) for their support.

\bibliography{ref}

\end{document}